\documentclass[a4paper,12pt]{article}
\usepackage{jinstpub}

\usepackage{lineno}
\usepackage{svg}
\usepackage{siunitx}
\usepackage{xspace}
\usepackage{hyperref}
\usepackage{url}
\usepackage{tabularray}

\newcommand{\Stot}{\ensuremath{\mathcal{S}}\xspace}
\newcommand{\Sice}{\ensuremath{\mathcal{S}_\mathrm{ice}}\xspace}
\newcommand{\Sgal}{\ensuremath{\mathcal{S}_\mathrm{Gal}}\xspace}
\newcommand{\Sele}{\ensuremath{\mathcal{S}_\mathrm{ele}}\xspace}
\newcommand{\Teff}{\ensuremath{T_\mathrm{eff}}\xspace}
\newcommand{\avg}[1]{\ensuremath{\langle #1 \rangle}\xspace}
\newcommand{\tc}[1]{$#1$\,\textdegree C}
\newcommand{\tk}[1]{$#1$\,K}
\newcommand{\rVEL}{\ensuremath{\mathcal{H}_r}\xspace}
\newcommand{\VEL}{\ensuremath{\mathcal{H}}\xspace}

\title{%
\hfill
  \begin{minipage}[l]{0.78\textwidth}
    A data-driven absolute amplitude calibration for the
    Radio Neutrino Observatory Greenland (RNO-G)
  \end{minipage}\hfill
    \begin{minipage}[r]{0.22\textwidth}
    \includegraphics[width=\linewidth]{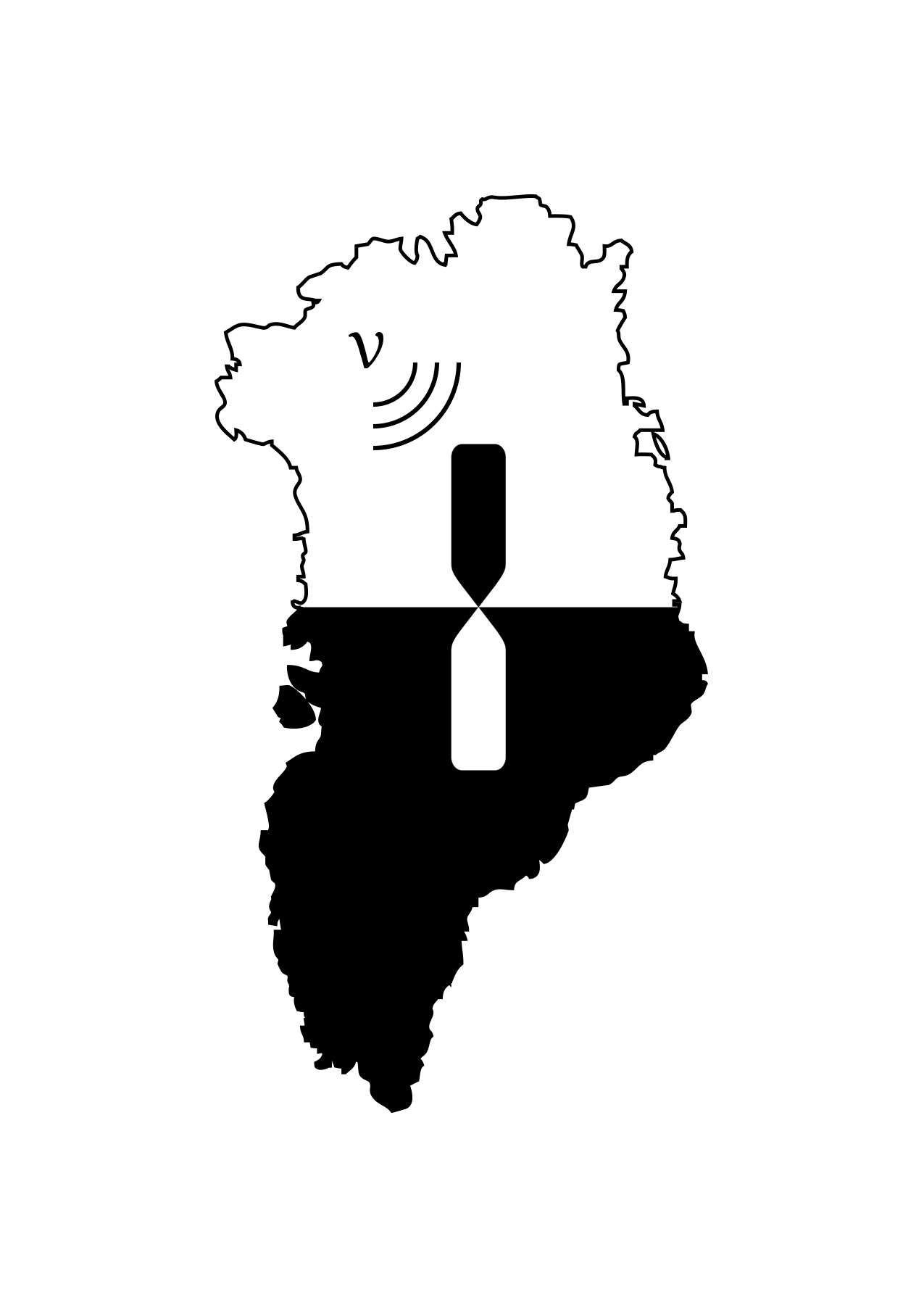}
  \end{minipage}
}

\collaboration{RNO-G Collaboration\\[6pt]}

\author[1]{S.~Agarwal}
\author[2]{J.~A.~Aguilar}
\author[3]{N.~Alden}
\author[1]{S.~Ali}
\author[4]{P.~Allison}
\author[5,6]{M.~Betts}
\author[1]{D.~Besson}
\author[7]{A.~Bishop}
\author[8]{O.~Botner}
\author[3]{M.~Boucher}
\author[9]{S.~Bouma}
\author[10,11]{S.~Buitink}
\author[2]{R.~Camphyn}
\author[7]{J.~Chan}
\author[2]{S.~Chiche}
\author[12]{B.~A.~Clark}
\author[1]{K.~Couberly}
\author[1]{D.~Dakroub}
\author[13]{K.~D.~de Vries}
\author[3]{C.~Deaconu}
\author[14]{P.~Giri}
\author[8,15]{C.~Glaser}
\author[4]{H.~Gui}
\author[8]{A.~Hallgren}
\author[9]{S.~Hallmann}
\author[16]{J.~C.~Hanson}
\author[17]{K.~Helbing}
\author[5,18]{B.~Hendricks}
\author[19,9]{J.~Henrichs}
\author[8]{N.~Heyer}
\author[1]{C.~Hornhuber}
\author[19]{E.~Huesca Santiago}
\author[4]{K.~Hughes}
\author[19,9]{A.~Jaitly}
\author[7]{A.~Karle}
\author[7]{J.~L.~Kelley}
\author[9]{C.~Kopper}
\author[2,13]{M.~Korntheuer}
\author[19,20]{M.~Kowalski}
\author[14]{I.~Kravchenko}
\author[5,18]{R.~Krebs}
\author[7]{M.~Kugelmeier}
\author[5,18]{D.~Kullgren}
\author[2]{A.~Kyriacou}
\author[9,19]{C.~Lagunas Gualda}
\author[9]{R.~Lahmann}
\author[14]{C.-H. Liu}
\author[4]{Y.~Liu}
\author[21]{M.~J.~Marsee}
\author[1]{C.~McLennan}
\author[1]{R.~Moser}
\author[11]{K.~Mulrey}
\author[7]{M.~Muzio}
\author[19,9]{A.~Nelles}
\author[22]{A.~Novikov}
\author[4]{A.~Nozdrina}
\author[3]{E.~Oberla}
\author[22]{N.~Punsuebsay}
\author[8]{M.~Ravn}
\author[15,23]{R.~Reimann}
\author[4]{Z.~Riesen}
\author[17]{A.~Rifaie}
\author[24]{D.~Ryckbosch}
\author[15,23]{Y.~Schaper}
\author[2,15]{F.~Schl{\"u}ter}
\author[13,25]{O.~Scholten}
\author[9]{P.~Schriefer}
\author[22]{D.~Seckel}
\author[1]{M.~F.~H.~Seikh}
\author[19,9]{Z.~S.~Selcuk}
\author[2]{G.~Singh}
\author[24]{J.~Stachurska}
\author[13]{J.~Stoffels}
\author[2]{S.~Toscano}
\author[7]{D.~Tosi}
\author[6]{J.~Tutt}
\author[13]{N.~van Eijndhoven}
\author[3]{A.~G.~Vieregg}
\author[12]{A.~Vijai}
\author[19,9]{H.~Warnhofer}
\author[15,23]{N.~Weitkemper}
\author[5,6]{C.~Welling}
\author[21]{D.~R.~Williams}
\author[19]{P.~Windischhofer}
\author[5,6]{S.~Wissel}
\author[9]{A.~Zink}
\affiliation[1]{University of Kansas, Dept.\ of Physics and Astronomy, Lawrence, KS 66045, USA}
\affiliation[2]{Universit\'e Libre de Bruxelles, Science Faculty CP230, B-1050 Brussels, Belgium}
\affiliation[3]{Dept.\ of Physics, Dept.\ of Astronomy \& Astrophysics, Enrico Fermi Inst., Kavli Inst.\ for Cosmological Physics, University of Chicago, Chicago, IL 60637, USA}
\affiliation[4]{Dept.\ of Physics, Center for Cosmology and AstroParticle Physics, Ohio State University, Columbus, OH 43210, USA}
\affiliation[5]{Dept.\ of Physics, Pennsylvania State University, University Park, PA 16802, USA}
\affiliation[6]{Dept.\ of Astronomy and Astrophysics, Pennsylvania State University, University Park, PA 16802, USA}
\affiliation[7]{Wisconsin IceCube Particle Astrophysics Center (WIPAC) and Dept.\ of Physics, University of Wisconsin-Madison, Madison, WI 53703,  USA}
\affiliation[8]{Uppsala University, Dept.\ of Physics and Astronomy, Uppsala, SE-752 37, Sweden}
\affiliation[9]{Erlangen Centre for Astroparticle Physics (ECAP), Friedrich-Alexander-University Erlangen-N\"urnberg, 91058 Erlangen, Germany}
\affiliation[10]{Vrije Universiteit Brussel, Astrophysical Institute, Pleinlaan 2, 1050 Brussels, Belgium}
\affiliation[11]{Dept.\ of Astrophysics/IMAPP, Radboud University, PO Box 9010, 6500 GL, The Netherlands}
\affiliation[12]{Dept.\ of Physics, University of Maryland, College Park, MD 20742, USA}
\affiliation[13]{Vrije Universiteit Brussel, Dienst ELEM, B-1050 Brussels, Belgium}
\affiliation[14]{Dept.\ of Physics and Astronomy, Univ.\ of Nebraska-Lincoln, NE, 68588, USA}
\affiliation[15]{Dept.\ of Physics, TU Dortmund University, Dortmund, Germany}
\affiliation[16]{Whittier College, Whittier, CA 90602, USA}
\affiliation[17]{Dept. of Physics, University of Wuppertal  D-42119 Wuppertal, Germany}
\affiliation[18]{Center for Multimessenger Astrophysics, Inst.\ of Gravitation and the Cosmos, Pennsylvania State University, University Park, PA 16802, USA}
\affiliation[19]{Deutsches Elektronen-Synchrotron DESY, Platanenallee 6, 15738 Zeuthen, Germany}
\affiliation[20]{Institut f\"ur Physik, Humboldt-Universit\"at zu Berlin, 12489 Berlin, Germany}
\affiliation[21]{Dept.\ of Physics and Astronomy, University of Alabama, Tuscaloosa, AL 35487, USA}
\affiliation[22]{Dept.\ of Physics and Astronomy, University of Delaware, Newark, DE 19716, USA}
\affiliation[23]{Lamarr Institute for Machine Learning and Artificial Intelligence, Dortmund, Germany}
\affiliation[24]{Ghent University, Dept.\ of Physics and Astronomy, B-9000 Gent, Belgium}
\affiliation[25]{Kapteyn Institute, University of Groningen, PO Box 800, 9700 AV, The Netherlands}

\abstract{The Radio Neutrino Observatory Greenland (RNO-G) aims to detect ultra-high-energy neutrinos (E $\geq 100\,$PeV) through the nanosecond-long coherent radio pulses produced by the Askaryan effect when neutrino-induced particle cascades develop in the Greenlandic ice sheet. 
Precise calibration of the analog signal chain of each antenna channel is essential for accurate reconstruction of neutrino-induced events and their physical properties.
In this work, we present a data-driven calibration method based on measurement of the ambient radio background. By combining first-principles simulations of the physical noise contributions with a parametric model of the detector response, we determine the absolute gain of each antenna channel. The resulting calibration provides a season-averaged gain for every channel with an accuracy of approximately 0.5\,dB (better than 5\%), while the investigated systematic uncertainties associated with the noise signal-model were found to be negligible.
This work represents the first application of an ambient-noise-based calibration that utilizes a multi-component model for the ambient noise, originally developed for radio air-shower experiments, to an in-ice radio neutrino detector.}


\emailAdd{authors@rno-g.org, ruben.camphyn@ulb.be, felix.schlueter@tu-dortmund.de}
\begin{document}
\maketitle

\section{Introduction}

The Radio Neutrino Observatory in Greenland (RNO-G) utilizes a network of in-ice radio antennas to monitor a large volume of ice for the detection of ultra-high-energy neutrinos (E $\geq 100\,$PeV). Neutrinos are detected through nanosecond-long radio pulses produced by the cascade of secondary particles in neutrino-nucleus collisions in the ice via the Askaryan effect \cite{Askaryan:1961pfb}. The attenuation length of radio waves in cold glacial ice, $\mathcal{O}(\mathrm{km})$, allows for the instrumentation of a vast volume with shallow detector units. This makes the in-ice radio detection technique particularly well suited for the detection of ultra-high-energy neutrinos which, given their expected low flux, require large effective detection volumes.

While the radio detection technique has been successfully applied to measure cosmic rays from extensive air showers \cite{LOFAR_2013, PierreAuger:2014ldh, Codalema_2017}, a detection of neutrinos with in-ice radio antennas has yet to be accomplished. Recently, the ARA collaboration announced the detection of radio pulses from in-ice particle cascades induced by cosmic-ray air showers hitting the high-altitude ice sheet~\cite{ARA:2025qit}. Similar work is ongoing in RNO-G~\cite{RNO-G:2025tjo} and a first detection is expected to be reported in the near future.

In order to accurately derive the emitted electric field of astrophysical signals from the detected voltages, an accurate calibration of the analog signal chain of each antenna channel is necessary. Throughout this work, the term analog signal chain refers to the entire instrumental response of an RNO-G antenna channel, excluding the directional radiation pattern of the antenna itself. Accurate knowledge of the analog signal chain is essential for efficient event selection, including the discrimination of cosmic-ray- and neutrino-induced signals from each other and from various noise sources, as well as for event reconstruction.  Furthermore, a well-characterized detector response is required to produce reliable detector simulations, for example for the training of machine-learning algorithms.

To record the antenna signals, RNO-G uses a complex network of radio-frequency (RF) and optical components: to reduce signal attenuation between the deep in-ice antennas and the data acquisition at the surface, optical fibers are used. While those components were individually tested and characterized in the lab, the in-situ system response can differ from this expectation due to the complex interplay of the various components. In particular, the optical fiber connections have been shown to be a source of deviations from the expected response: disconnecting and reconnecting the fibers can significantly change the system response, in particular the gain. While refinements of the in-lab calibration and installation procedures are ongoing, this motivates the development of a data-driven, in-situ calibration method to determine the absolute system amplification.

\begin{figure}
    \centering
    \includegraphics[width=\linewidth]{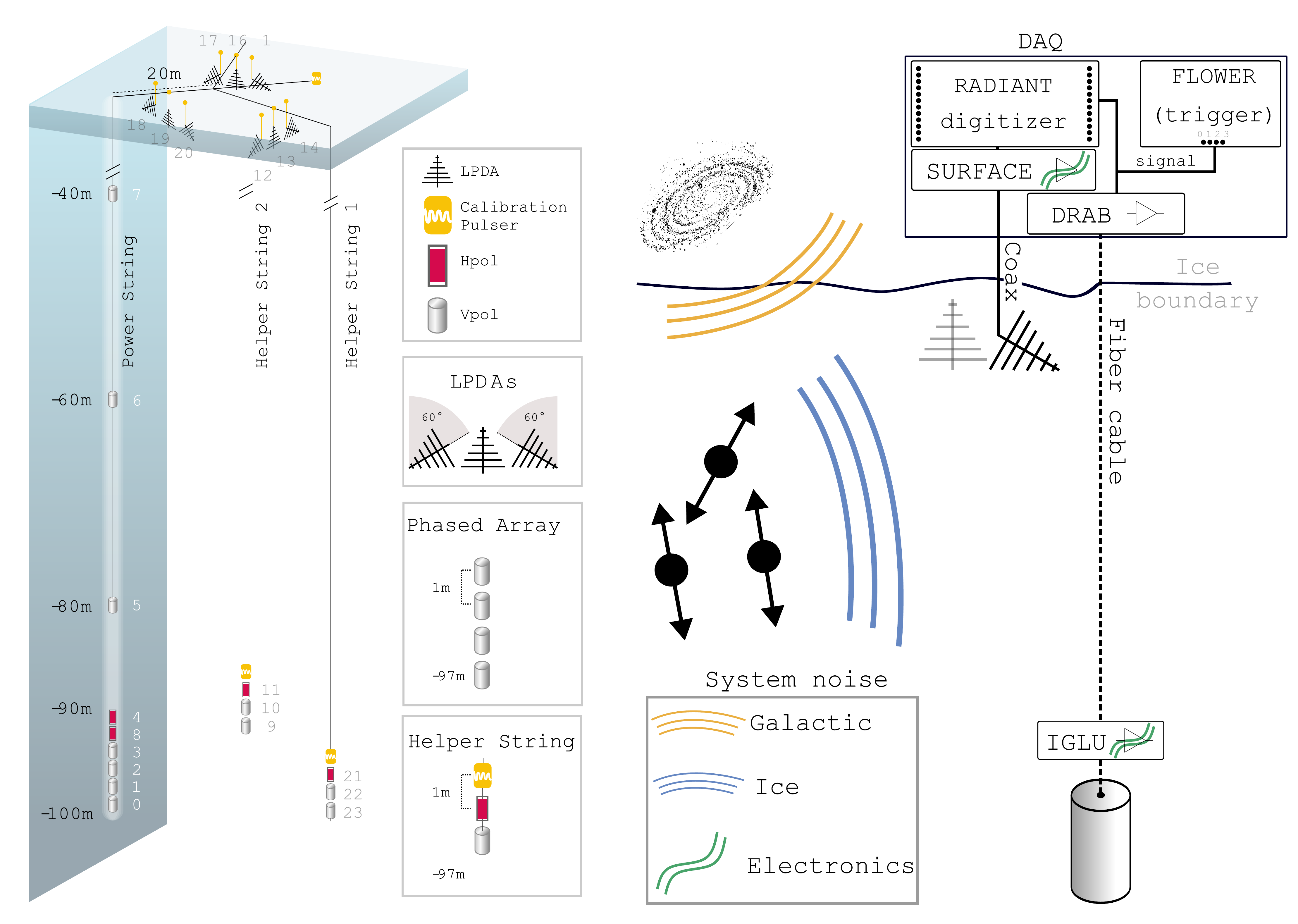}
    \caption{\textit{Left}: Schematics of the RNO-G station design. Nine antennas are part of the ``shallow component'' and installed in shallow trenches beneath the surface, and the remaining 15 antennas are part of the ``deep component'' and installed in three 100$\,$m deep boreholes. \textit{Right}: Schematic of a single deep and shallow signal chain. Indicated are the different noise sources and their point of entry in the signal chain.}
    \label{fig:station}
\end{figure}

To this end, we model the ambient background radiation originating from the ice and the sky, the internal electronic noise, and the analog signal chain to simulate the measured noise spectra, which are then compared with the data. Similar approaches have been developed for air-shower radio detectors such as LOFAR \cite{Mulrey:2019vtz} and AERA \cite{PierreAuger:2023hun}. The in-ice radio neutrino detector ARA \cite{ARA_sensitivity} utilizes a data driven calibration where the signal-chain response is inferred from ambient noise measurements using a single-component flat-temperature (not frequency dependent) noise model. 

The method described here is only sensitive to the analog frequency-dependent amplitude response of the instrument. The direction-dependent antenna sensitivity and the channel's phase response are not calibrated. For a description of the directional antenna sensitivity, we rely on simulations, while the phase response of the analog signal chain is taken from lab measurements.

This paper is structured as follows: After introducing the detector and general concepts in Section~\ref{sec:concepts}, we present the simulation in Section~\ref{sec:sim}. Results for all stations and antenna channels are shown in Section~\ref{sec:results}, while the associated calibration uncertainties are discussed in Section~\ref{sec:uncertainties}. We draw conclusions in Section~\ref{sec:conclusion}.

\section{Fundamentals and Concepts}
\label{sec:concepts}
First, the experimental design of RNO-G is described in Section \ref{sec:detector}. In Section \ref{sec:concept}, the calibration concept is presented. Details about the data are presented in Section \ref{sec:data}, with a focus on system response in Section \ref{sec:response}.
\subsection{Detector}
\label{sec:detector}
The Radio Neutrino Observatory in Greenland is located near the apex of the Greenland ice sheet on top of $\sim\SI{3200}{m}$ glacial ice. It is co-located with the NSF-operated Summit Station, which provides a base for deployment and maintenance activities and hosts equipment for data collection and communication. RNO-G is funded to feature a 35-station array covering an area of roughly 50\,km$^2$ \cite{RNO-G:2020rmc}. Prior to the 2026 deployment season, RNO-G consisted of eight independent and, mostly, autonomously operating radio-antenna stations. Each station comprises 24 antennas installed at different depths within the ice. Nine log-periodic dipole antennas (LPDA) are deployed in trenches just beneath the surface and referred to as the ``shallow'' component\footnote{At deployment, the antennas are at a depth of $\sim\SI{1}{m}$. However, due to snow accumulation the depth increases by less than \SI{1}{m} per year.}. The remaining 15 antennas, 11 vertically polarized fat-dipole antennas (VPols) and 4 horizontally polarized quad-slotted cylinder antennas (HPols), are installed in three \SI{100}{m} deep boreholes and referred to as the ``deep'' component. A sketch of the station design is shown in Fig.~\ref{fig:station} (\textit{left}). 
The four VPol-channels at the bottom of the ``power string'' are operated as a phased array to provide a low-threshold neutrino trigger \cite{Allison:2018ynt}.

The signals of all 24 antennas are digitized and recorded with a common data acquisition system (DAQ) situated near the surface.
The DAQ uses two digitizer boards, the 24-channel RADIANT board and the 4-channel FLOWER board. The RADIANT digitizes the signals of each antenna with 12-bit resolution at a sampling rate of up to \SI{3.2}{GHz}
using a custom switched-capacitor array chip called LAB4D \cite{Roberts:2018xyf} for each channel. Each waveform has 2048 samples, corresponding to a few hundred nanoseconds.
The FLOWER digitizes the signals of the four phased-array channels (0--3) at 472\,MS/s and 8-bit resolution with a dynamic gain equalizer.
While the full-band RADIANT-digitized waveforms are read out and stored for each trigger, FLOWER-digitized waveforms are used for triggering and are not stored during normal data taking.

The design of the analog signal chain is different between the deep and shallow components. To reduce attenuation, the deep component features a custom RF-over-fiber (RFoF) design transmitting the RF signals from the antennas in the boreholes to the DAQ at the surface via optical fibers. In contrast, the shallow component uses coaxial cables for signal transmission. A detailed description of the entire instrumental design of both components and digitizer boards is given in \cite{RNO-G:2024esr}. Here, we summarize the most important facts relevant for the performed calibration. A sketch of the two signal chains is depicted in Fig.\ \ref{fig:station} (\textit{right}).

Each deep antenna is connected via a short coaxial cable to a single-channel low-noise amplifier, the ``In-ice Gain with Low-noise Unit'' (IGLU), deployed alongside the antenna in the borehole. The IGLU combines a 50\,$\Omega$-matched low-noise amplifier with an amplitude-modulated RFoF transmitter. The optical signals are carried to the surface through a multi-channel fiber bundle and received by four 4-channel ``Downhole Receiver and Amplifier Boards'' (DRABs), which convert them back to electrical signals and apply further amplification and filtering. Each DRAB output is connected to a RADIANT digitizer channel via a short coaxial cable; for the four phased-array channels (0--3), a symmetric splitter additionally routes the signal to the FLOWER trigger board.

The surface LPDA antennas are connected via coaxial cables of about 11\,m total
length to one of two 5-channel SURFACE boards, which provide amplification and filtering before the signals are passed to the RADIANT.

Of the eight stations in operation before the 2026 field season\footnote{Four additional stations were deployed in the 2026 field season.}, three were first deployed in 2021 (stations 11, 21, and 22), four in 2022 (stations 12, 13, 23, and 24) and one in 2024 (station 14). The four stations deployed in 2022 feature modified IGLU and DRAB boards.  
In the context of this paper, all stations except station~14 are calibrated using data recorded in 2022, 2023 and 2024, prior to the replacement of the DAQ in several stations during the 2024 field season. Station~14 is equipped with a newer revision of the RADIANT digitizer board (v3), for which full-chain lab measurements of the system response were not available during the development of this method. Such measurements have since been performed, and the calibration of station~14 will be performed analogously in the future.
\subsection{Methodology}
\label{sec:concept}
The calibration method uses thermal noise as a standard candle and tunes a model of the system response to match the simulation of this noise to data. This requires three ingredients: I) a model for the standard-candle noise \Stot, II) a parametric model of the system response $\mathcal{R}$, and III) a clean data sample of this background noise $\mathcal{D}$, free of contamination from unmodeled noise sources such as anthropogenic RF interference or transient wind-related backgrounds as well as physics events such as cosmic-ray- or neutrino-induced radio signals.

To describe the entire noise signal recorded by an RNO-G antenna channel \Stot, we identified three independent contributions: I) the thermal radiation of the ambient glacial ice \Sice in which the antennas are embedded, II), the radiation of the Galaxy \Sgal, and III) the electronic noise \Sele from the RF components of the antenna channels. In contrast to air-shower detectors, where radiation from our Galaxy is the dominant background, the dominant ambient contribution to the recorded signals in in-ice neutrino detectors, which also operate at higher frequencies, is the thermal radiation of the ice.

Both the thermal radiation of the ice and that of the Galaxy first pass through the antenna before they are amplified by the signal chain, which is described by $\mathcal{R}$. Hence, in order to correctly simulate the antenna-channel signals, the direction-dependent radiation ``pattern'' of the antenna needs to be taken into account. We describe the antenna pattern via the realized vector effective length $\rVEL(f, \phi, \theta)$, which relates the incoming electric field $E$ to a voltage at a 50\,$\Omega$-matched signal chain (receiver) connected to the antenna terminal $U=\rVEL\,E$ (we omit the notation for vectors here for \rVEL and $E$). 

The open-source framework NuRadio \cite{Glaser:2019cws,Glaser:2019rxw} is used to simulate the background, model the instrumental response of RNO-G antenna channels, and process RNO-G data. For more details see Appendix~\ref{app:code}. In the following subsections, we describe the construction of all three ingredients $\mathcal{S}$, $\mathcal{R}$, and $\mathcal{D}$.

\subsection{Data}
\label{sec:data}
To obtain a clean data set of the noise without any significant contribution of unmodeled sources such as wind-induced noise or anthropogenic RF interference (RFI), we only use data from a \SI{0.1}{Hz} forced software trigger, thus excluding possible cosmic-ray or neutrino events.
Furthermore, we only select data from 2-hour data-taking ``runs'' with a maximum total trigger rate of \SI{2}{Hz} (nominal trigger rates vary between stations and in time but thresholds are tuned to a target rate of $\sim\SI{1}{Hz}$).

To further suppress RFI, we apply a continuous-wave (CW) removal algorithm which detects monochromatic continuous waves by their amplitude in the frequency domain, fits a sine wave of that frequency in the time domain, and subtracts it. This procedure has already been used in other radio experiments \cite{PierreAuger:2015hbf,ANITA:2018vwl}.

Where possible, a non-linear conversion between recorded ADC counts and voltage is applied. This ``voltage calibration'', based on the ADC response from a scan of the pedestal voltage, describes minor deviations from linearity in the response of the digitizer\footnote{These scans are taken once a day from 2023 onward. Hence, for data in 2022 a linear conversion, $1\,\mathrm{ADU} = 2.5\,\mathrm{V} / (2^{12}-1)$, is used.}. All waveforms are filtered from 100\,MHz to 700\,MHz with a digital 10th-order Butterworth filter. Our data processing also includes a baseline correction and the removal of a digitizer artifact in which blocks of 64 samples are occasionally recorded out of order.

\begin{figure}
    \centering
    \includegraphics[width=\linewidth]{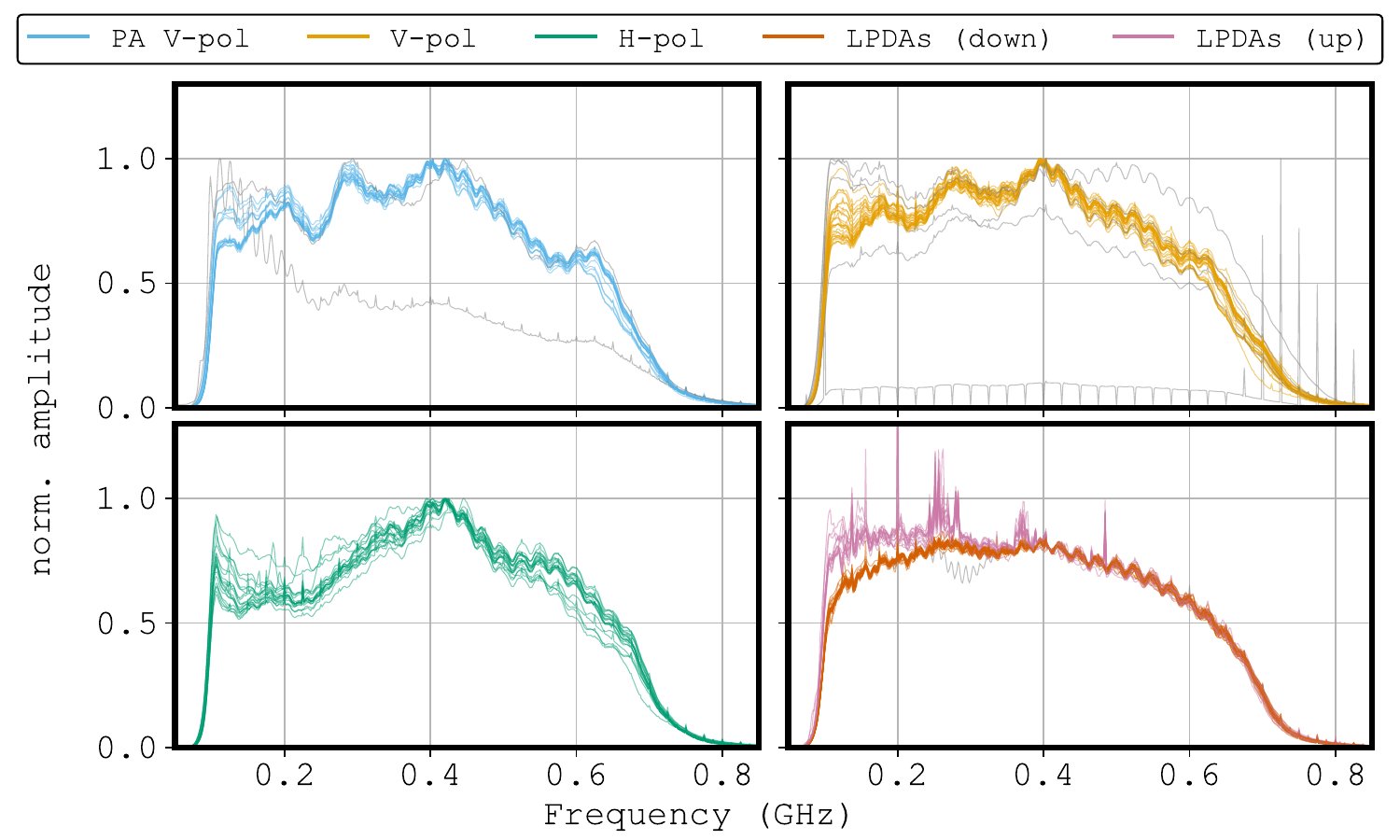}
    \caption{Normalized, averaged spectral amplitudes for the different channel categories. Each spectrum corresponds to one channel in one station. Visibly broken channels are shown in light gray, see details in the text.}
    \label{fig:response}
\end{figure}

Fig.~\ref{fig:response} shows the normalized spectra for data recorded in 2023 across five stations. Stations 21 and 22 are excluded from this plot. Station 21 is the closest to Summit Station, and despite the RF filtering its spectra exhibit several narrow-band RFI features; the station is nonetheless included in the calibration procedure, as this RFI has no significant effect on our calibration. Station 22 began malfunctioning in 2023 and is therefore excluded from the calibration for both 2023 and 2024\footnote{Its DAQ was replaced in 2024, and the station is operational again.}. We also identified malfunctioning channels in the remaining stations (shown as light gray lines) and excluded these from the evaluation of this calibration as well.

The channels are sorted into five types: phased-array VPols (which differ from the other VPols by an additional signal splitter between the DRAB and the RADIANT), other VPols, HPols, and upward- and downward-facing LPDAs. Each type exhibits distinct features that can be attributed either to a different instrumental response (e.g.\ VPols vs.\ HPols, or deep vs.\ shallow channels) or to different levels of ambient noise (e.g.\ upward- vs.\ downward-facing LPDAs). Differences between shallower and deeper VPols are minimal. Within each group, the normalized spectra agree well, particularly at intermediate and higher frequencies. At low frequencies ($\lesssim 150\,$MHz) a larger scatter is seen in the deep antennas. A channel-dependent impedance mismatch between the antennas and IGLUs could, at least partially, explain this scatter. We elaborate on this in Appendix~\ref{app:impedance}. A clear excess in the upward-facing LPDAs over the downward-facing LPDAs at low frequencies is visible. Furthermore, the upward-facing LPDA spectra exhibit prominent narrow-band RFI, which persists even after RF cleaning. 

The observed similarity within channels of one type (modulo the aforementioned differences) motivates describing the analog signal-chain response of the phased-array VPols, the other deep channels, and the surface channels through a shared ``response template'' that captures the phase and relative amplitude response as a function of frequency. The only differing factor between channels within the same channel type is the absolute gain.

\subsection{System response}
\label{sec:response}

This calibration relies on ``template'' responses which describe the phase and relative frequency response of a channel. The templates are obtained from lab measurements with a picosecond pulse generator. In our setup, the picosecond pulse generator replaces the antenna and connects via the respective coaxial cable to either an IGLU or SURFACE board. The signals are recorded with the full DAQ including the digitizer board. By averaging over hundreds of individual pulses, reducing the effect of thermal fluctuations, we obtain a measurement for the system response in the time domain $R(t)$. It should be noted that this is equivalent to a Vector Network Analyzer (VNA) measurement of the $S_{21}$ scattering parameter in the frequency domain. However, following this approach, it is possible to include the analog response of the digitizer board, which is not possible with a VNA measurement. We also found that concatenating the $S_{21}$ parameters of individual components in the signal chain, which we refer to as the \emph{nominal} (uncalibrated) response in the following, does not provide a satisfactory description of our data. Therefore, the present analysis relies on full-chain time-domain measurements, of which only a limited number are available.

Simultaneously with the DAQ recording, the signals of the picosecond pulse generator are measured by an oscilloscope. The frequency-dependent amplitude spectrum of the impulsive signals is found to follow $w_p(f) \equiv 1 - \frac{0.352}{\mathrm{GHz}} \left(f - f_0\right) $ with $f_0 = 0.1\,$GHz within our bandwidth ($<800$\,MHz). This response of the pulse generator has to be unfolded from the measured response of the analog signal chain. Hence, we divide the measured channel responses by $w_p(f)$ (cf.\ Eq.\ \ref{eq:response}).

\begin{figure}
    \centering
    \includegraphics[width=\linewidth]{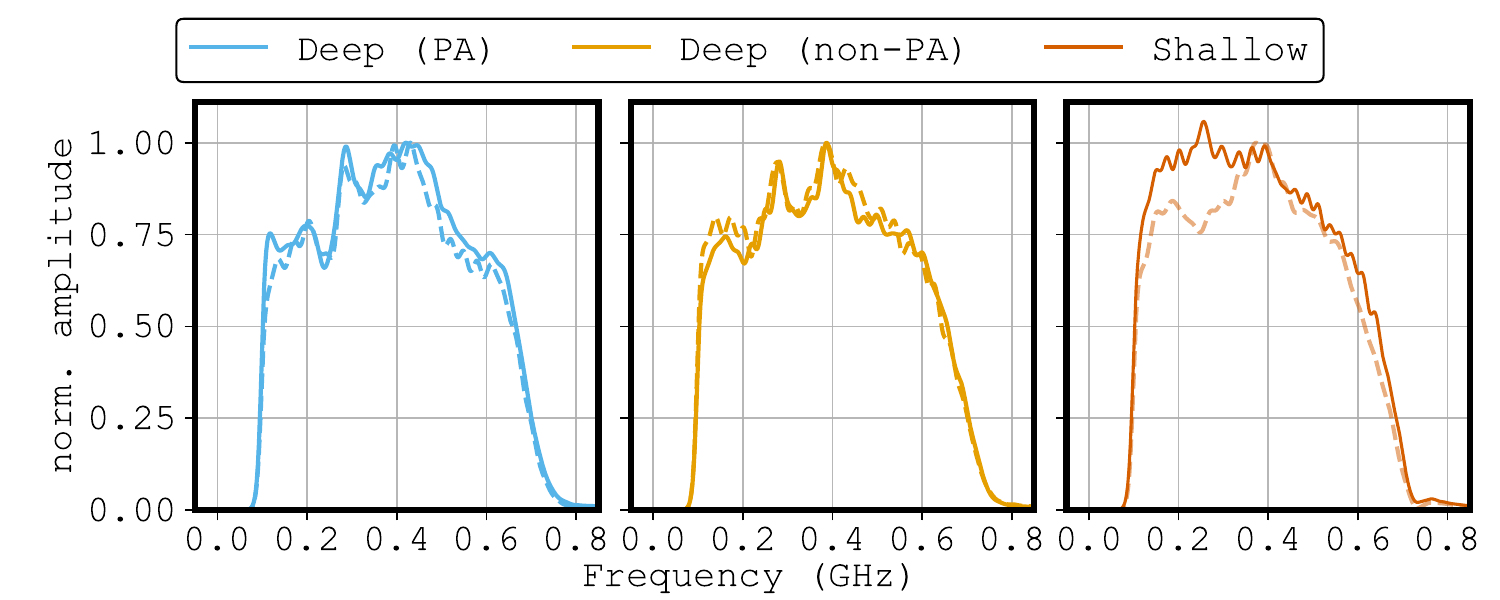}
    \caption{Compilation of the system response templates, grouped by response chain type. Two separate measurements were made for three different signal chain configurations (deep phased-array channel, deep channel, shallow channel), yielding a total of six possible templates. The measurements are differentiated by their linestyles (solid vs dashed) while the color indicates the signal chain configuration. The two left panels are normalized to $1$, the right panel is normalized such that the value at 400 MHz is set to 1 to better visualize the differences between the two measurements.}
    \label{fig:template_responses}
\end{figure}

Without the antenna, we distinguish between three different response chains: a deep phased-array channel, a deep (non-phased-array) channel, and a shallow channel. Fig.~\ref{fig:template_responses} shows $\hat{\mathcal{R}}_i(f)$ for two measurements for each of the three different signal chains. Modifications to the IGLUs and DRABs (cf.~Sec.~\ref{sec:detector}) between the stations deployed in 2022 and 2023 are not reflected in these templates as dedicated measurements for the components deployed in 2022 are missing. 

While the normalized spectra of the deep channels exhibit a strong scatter at lower frequencies (cf.~Fig.~\ref{fig:response}), the different measurements of $\hat{\mathcal{R}}_i(f)$ do not reflect that scatter. We found that an additional weighting function $w_s(f) = 1 + \frac{s}{\mathrm{GHz}} (f - f'_{0})$ with $f'_{0} = 400\,$MHz and $s$ a free fit parameter, improves the description of the data.

Combining all this, we obtain a parametric model of the signal chain with the absolute gain as a free parameter:
\begin{equation}
    \label{eq:response}
    \begin{gathered}
    \mathcal{R}_i(f, G) = G \cdot \hat{\mathcal{R}}_i(f), \quad
    \hat{\mathcal{R}}_i(f) = \mathcal{R}'_i(f) ~ / ~ \max\{\mathcal{R}'_i(f)\} \\[6pt]
    \mathcal{R}'_i(f) = \frac{w_s(f) ~ \mathrm{RFFT}\{R_i(t)\}}{w_p(f)}
    \end{gathered}
\end{equation}
where $\hat{\mathcal{R}}_i(f)$ is normalized such that $\mathrm{max}(\hat{\mathcal{R}}_i(f)) \equiv 1$. Here the index $i$ denotes different lab measurements. RFFT denotes the real fast Fourier transform, which follows the convention in \cite{Glaser:2019cws}. Since the slope $s$ is fitted, the normalization is repeated in every step of the minimization process. 

For each channel, the best-fitting template $\mathcal{R}_i(f)$, based on a $\chi^2$ goodness-of-fit criterion, is used. We intentionally did not restrict the choice of template to the corresponding channel type in order to test whether the correct association would emerge naturally. For the shallow channels, a single shallow response was selected exclusively (cf.\ Fig.\ \ref{fig:template_responses}, \textit{right}). Among the deep channels, only one non-phased-array VPol channel was best described by a phased-array template, whereas all other VPol channels were best fitted by templates corresponding to their respective channel types. The HPol channels preferentially selected phased-array templates, although these are, strictly speaking, not appropriate for HPol channels. 

\section{Simulating standard-candle background}
\label{sec:sim}
The three contributions to the background noise in RNO-G are illustrated in the sketch in Fig.~\ref{fig:station} (\textit{right}). The ice and Galactic radiation constitute the external noise contributions and are described by a frequency- and direction-dependent brightness temperature. Following that, the electric field at the antenna is calculated using the spectral radiance $B_\nu$, the power per unit area, steradian, and frequency $\nu$ according to the Rayleigh--Jeans law

\begin{equation}
    \label{eq:radiance}
    B_\nu^0 (T) = \frac{2 k_B \, \nu^2\,T}{c_0^2}
\end{equation}

where $T$ is the (apparent) brightness temperature of the emission at the antenna, $k_B$ the Boltzmann constant, and $c_0$ the speed of light in vacuum. As we describe the electromagnetic radiation in the dielectric medium ice with the refractive index $n_\mathrm{ice} \approx 1.3$--$1.8$ (depending on the depth), the spectral radiance becomes $B^n_\nu(T)= n^2 B^0_\nu(T)$, see p.\ 331 of \cite{Clausius1879}. The spectral radiance is converted to a flux density (Poynting vector $S_\nu = B_\nu \delta\Omega \delta \nu$) by multiplying it by the solid angle $\delta\Omega$ and frequency bin $\delta \nu$. The time integrated power of a plane wave at the antenna then gives an electric-field amplitude of

\begin{equation}
    \label{eq:efield}
    E = \sqrt{S_\nu ~ Z} = \sqrt{n \frac{B^0_\nu(T) \delta \nu \,\delta\Omega}{c_0\,\epsilon_0}}.
\end{equation}
using the impedance of the ice $Z = Z_0 / n = 1 / \left(c_0 \, n \, \epsilon_0\right)$. Note that the electric-field amplitude scales with $\nu$, which causes a white (i.e. constant-temperature) spectrum at the antenna to induce more power at higher frequencies. As mentioned in Sec.~\ref{sec:concept}, to simulate the signal at the antenna terminal, the electric field is converted using the antennas' realized vector effective length which typically falls off as $\nu^{-1}$ resulting in a mostly flat spectrum at the antenna terminal (for a flat input spectrum).

The electronic noise is internal and described by a frequency-dependent temperature. The voltage amplitude it induces is calculated according to Johnson--Nyquist via the power it delivers into the signal chain
\begin{equation}
    \label{eq:ele_noise}
    V_\mathrm{RMS} = \sqrt{T(\nu) \, R_L \, k_B \,\Delta\nu}
\end{equation}
with $R_L$ being the impedance of the signal chain (at the stage of injection) and $\Delta\nu$ the (effective) bandwidth. For the purpose of this calculation, we assume a purely resistive signal chain at exactly \SI{50}{\ohm}.

\subsection{Thermal radiation from the ice}
\label{thermal_ice}

The dominant external contribution is the thermal radiation of the glacial ice. It originates from the Brownian motion of the ice molecules and is characterized by the physical temperature $T_\mathrm{ice}(z)$ and the emissivity $\epsilon(z)$ of the ice. Both can be described as functions of depth $z$. The temperature profile has been measured with ice-core probes~\cite{grip}. Apart from seasonal variations in the uppermost few meters, the ice temperature is nearly constant at about \tc{-32} down to a depth of 2\,km and then increases gradually to approximately \tc{-10} at the bedrock below 3\,km. Within those first few meters the temperature varies seasonally between \tc{-40} in winter and \tc{-15} in summer~\cite{Giese_Hawley_2015}; we verified that varying the temperature in the upper \SI{20}{m} has a negligible effect. The emissivity equals the absorption coefficient, $\epsilon(z) = \alpha(z) = 1 / l_\mathrm{att}(z)$, with $l_\mathrm{att}(z)$ being the depth-dependent attenuation length. The total attenuation at the site has been measured by the RNO-G collaboration with radio echoes~\cite{faucris.276704036}. The attenuation length follows the temperature profile to first order: it is largest, i.e.\ the attenuation is weakest, at shallow depths and decreases towards the bedrock.

Because the refractive index of glacial ice changes over the uppermost $\sim$100\,m (the layer commonly referred to as \emph{firn}), radio waves do not propagate along straight lines, and the corresponding ray paths are determined numerically using the ray-tracing code \texttt{RadioPropa}~\cite{Oeyen:2021aco}. Rays are traced from the antenna for equidistant incidence angles until they reach the bedrock at $z_\mathrm{bottom} = 3027.6\,$m, including reflection at the air--ice boundary at $z = 0$ and the associated losses.

The effective temperature at the antenna, \Teff, is obtained by integrating the emission along these ray paths for given incidence directions $\theta$, accounting for re-absorption in the ice and for the contribution of the bedrock. Assuming that the ice is homogeneous in $x$ and $y$ and that its properties vary only with depth, $\Teff^z(\theta)$ is, for a given antenna depth $z$, a function of the zenith angle only. Applying the equation of radiative transfer to a single ray path $s(z)$ yields

\begin{align}
    T_\mathrm{eff}(\theta) &= T_\mathrm{rock} \, a(s_\mathrm{max}) + \int_0^{s_\mathrm{max}} a(z) \, \epsilon(z) \, T(z) \, ds \notag \\
             &\approx T_\mathrm{rock} \, e^{- s_\mathrm{bot} / \langle l_\mathrm{att} \rangle_\mathrm{bot}} + \sum_{i}^{i = \mathrm{bot} - 1} T(z_i)  \, \Delta s / l_\mathrm{att}(z_i) \, e^{-  s_i / \langle l_\mathrm{att} \rangle_i},
    \label{eq:teff}
\end{align}

\noindent with $a(s)= e^{-s(z) / \langle l_\mathrm{att} \rangle_z}$ the attenuation along the ray path;  $z_i = z(s_i)$ and $s_i = \sum_{j=0}^i \Delta s$. The first term represents the black-body radiation from the rock while the second term describes the propagation of the radiation including absorption and re-emission. Reflection losses at the air--ice boundary enter as an additional factor $\Theta^\theta(s_i)$

\begin{equation}
    \Theta^\theta(s_i) =     
    \begin{cases}
        1  & \text{for ray paths without reflection} \\
        |r_i|^2 & \text{for ray paths reflecting at the surface}
    \end{cases}
\end{equation}
with the Fresnel reflection coefficient $r_i$ calculated by \texttt{RadioPropa} in each term of Eq.~\ref{eq:teff}.

\begin{figure}
    \centering
    \includegraphics[width=\linewidth]{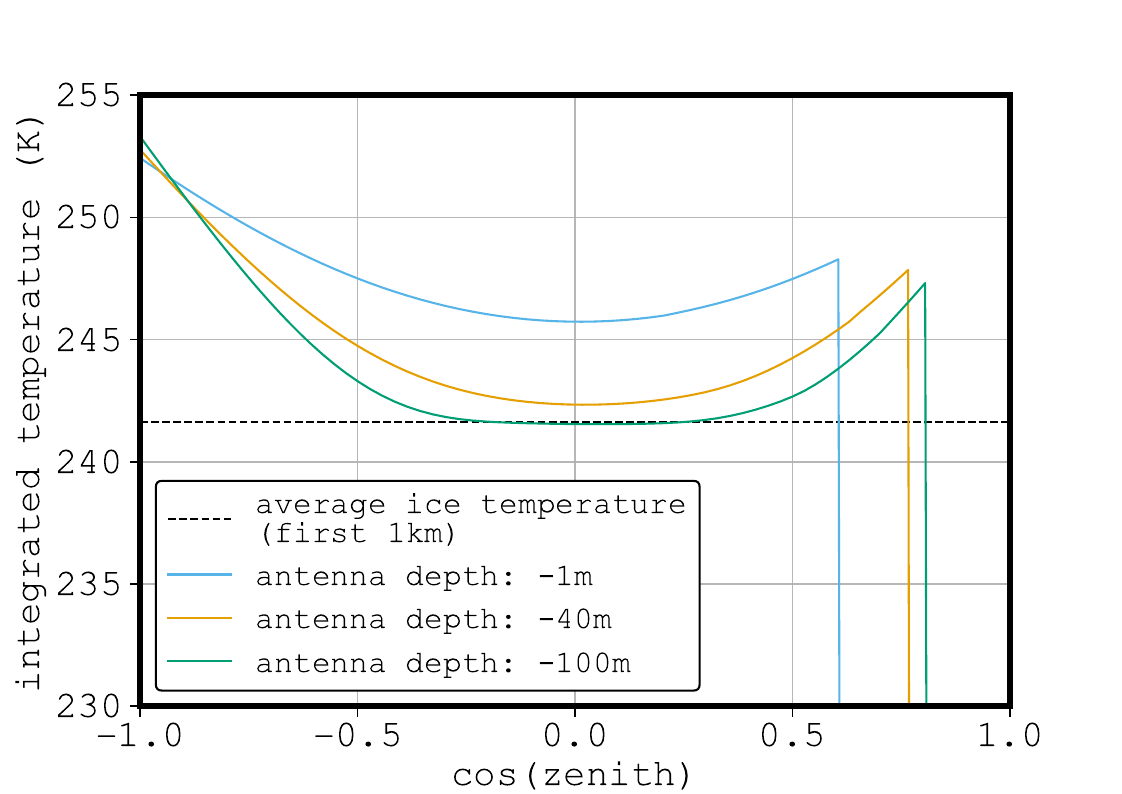}
    \caption{Effective temperature at the antenna as a function of the incidence angle for three different antenna depths.}
    \label{fig:eff_temp}
\end{figure}

Fig.~\ref{fig:eff_temp} shows \Teff as a function of the incidence angle for three antenna depths. For all depths the temperature is highest for directions pointing straight down towards the bedrock, where the ice is warmer. It decreases towards the horizon ($\cos(\theta) = 0$), increases slightly for upward-pointing directions, and then drops sharply. This drop marks the angle of total internal reflection at the air--ice boundary. As a result, the thermal emission received from behind the reflection point is strongly suppressed, limiting the volume of ice that effectively contributes to the measured signal. The position of the drop depends on the antenna depth and shifts to more upward-pointing directions for deeper antennas. Beyond the drop, the temperature does not fall to zero but instead approaches a plateau of approximately \tk{50}, with the exact value depending on the antenna depth (not shown). The absolute scale of the curves is set by the geometry of the ray trajectories. Rays reaching the shallowest antenna are refracted more, due to its position in the firn where the refractive index changes most rapidly; these rays therefore reach deeper, warmer ice over shorter path lengths.

\subsection{Thermal radiation from the sky}

The Galaxy, and in particular the Galactic Center, are bright radio sources, especially below 100\,MHz (below our calibration band) where they constitute the dominant background for air-shower experiments. Several models exist that describe this emission in terms of the sky's brightness temperature as a function of frequency and direction. These models are based on measurements from dedicated radio telescopes, which operate in narrow, experiment-specific frequency ranges and only cover parts of the full sky. Hence, these models interpolate and extrapolate between those measurements to arbitrary frequencies and positions in the sky. An extensive review of several available models and their differences can be found in \cite{Busken:2022mub}. For the purpose of this work, we use the Global Sky Model (GSM2008) \cite{de_Oliveira_Costa_2008}. Since the Galactic Center is not visible from the RNO-G site, the diurnal variation of the Galactic emission caused by the changing field of view is small: the relative strength of the emissions varies by $\pm 5\%$ between the daily minimum and maximum in GSM2008. Moreover, the fitted gain is mainly constrained by the higher frequencies, even for the upward-facing LPDAs for which the emission is dominant below 100\,MHz (see Fig.\ \ref{fig:results}). We therefore model the Galactic thermal emission as a 24-hour average, integrating over the entire visible sky.

The sky temperature is converted to an electric field using Eqs.~\ref{eq:radiance}\,\&\,\ref{eq:efield} at the air--ice boundary and propagated into the ice and to the antennas taking into account reflection, refraction, and attenuation effects. In particular, reflection and refraction affect the antenna signal while attenuation effects are subdominant for deep antennas. The contribution of this emission to the overall signal received by the RNO-G antennas is shown in Sec.~\ref{sec:results}, Fig.\ \ref{fig:results}. Uncertainties from the sky model are based on the model comparison study \cite{Busken:2022mub} and have a negligible effect on the final calibration (see Sec.~\ref{sec:signal_uncert}). We also investigated the effect of the quiet Sun using a model for its apparent brightness in the RNO-G frequency band from Fig.~5 in \cite{Zhang_2022} and found it to be negligible.

\begin{figure}
    \centering
    \includegraphics[width=0.7\linewidth]{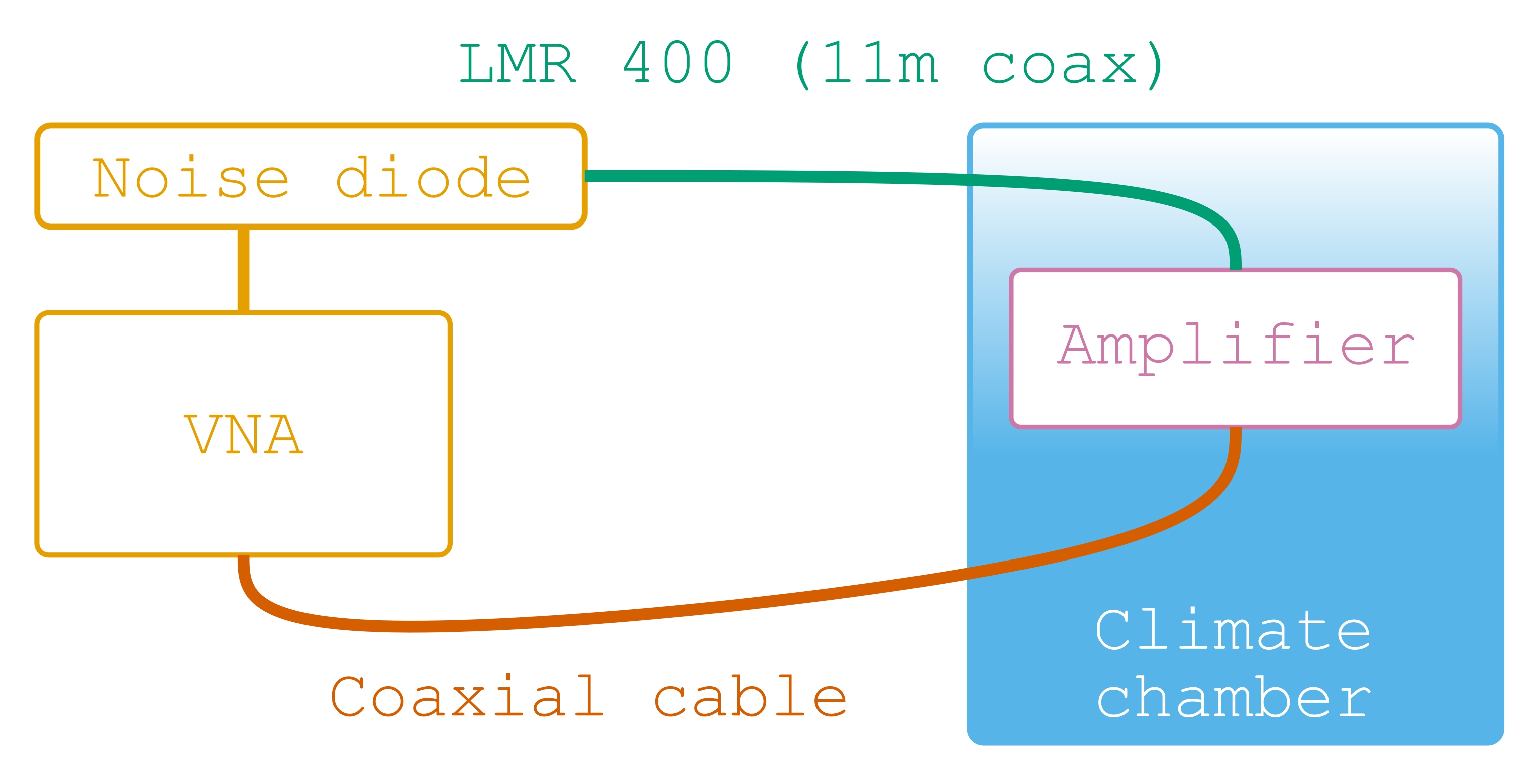}
    \caption{Sketch of the setup used to measure the noise figure for the deep and shallow signal chains in the lab.}
    \label{fig:noise_setup}
\end{figure}

\subsection{Electronic noise}
\label{sec:elec_noise}
The noise produced by the RNO-G electronics is measured in the lab using a climate chamber, a VNA, and a calibrated noise source. A sketch of the setup can be seen in Fig.~\ref{fig:noise_setup}. The noise source connects through an 11\,m LMR-400 coaxial cable (the same as deployed with the shallow antennas in RNO-G) to either a SURFACE or an IGLU placed inside the climate chamber. The climate chamber allows us to control the temperature down to \tc{-40}.

The SURFACE board connects through another coaxial cable to the VNA outside of the climate chamber. Noise measurements for the SURFACE amplifier were taken at \tc{0} as the SURFACE boards are located in the insulated DAQ box which in operation maintains higher temperatures than the surrounding ice.

The IGLU is connected to the VNA through a fiber, DRAB, and coaxial cable. For the deep component the measurement is performed at \tc{-40} reflecting the low temperatures in the boreholes. Details on how the noise temperature is derived from the VNA measurements for the different signal chains are given in Appendix~\ref{app:electronic_noise_figure}.

Fig.~\ref{fig:elec_noise} shows the measured noise temperatures for both signal chains. While the temperature in the borehole is expected to remain very stable throughout the year, the temperature in the DAQ box may change within a season following the ambient temperature and between seasons due to the changing depth of the DAQ box. Any resulting uncertainty in the magnitude of the electronic noise and thus our calibration is discussed in Sec.~\ref{sec:signal_uncert}.

To simulate individual noise spectra, using Eq.~\ref{eq:ele_noise}, the magnitude of the signal in the frequency domain is drawn according to a Rayleigh distribution with the scaling parameter $\sigma = V_\mathrm{RMS} ~ N / \sqrt{2\,N_\mathrm{band}}$ per frequency bin. $N$ is the total number of frequency bins in the simulated traces, $N_\mathrm{band}$ that of the selected frequency range. The phases are drawn uniformly from $[0, 2\pi)$.

\begin{figure}
    \centering
    \includegraphics[width=\linewidth]{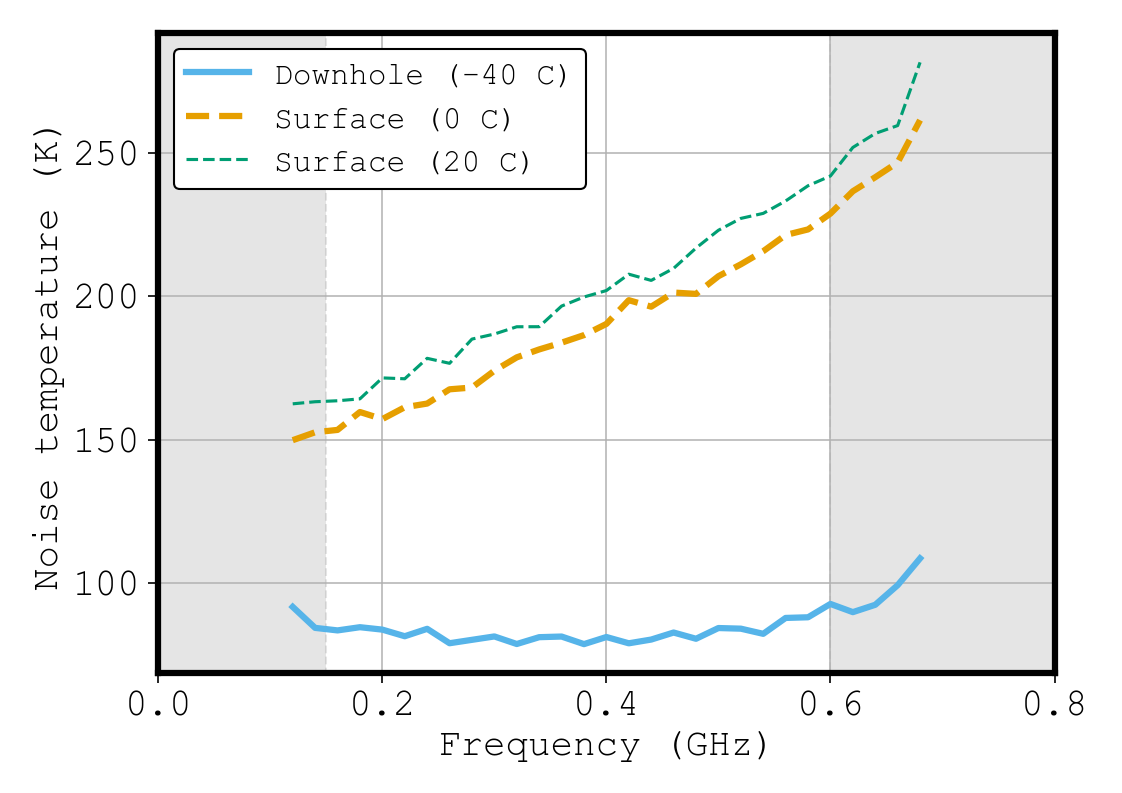}
    \caption{Noise temperature measurements taken for the downhole signal chain at \tc{-40} and the surface signal chain at both \tc{0} and \tc{20}.}
    \label{fig:elec_noise}
\end{figure}    

\section{Calibration Results}
\label{sec:results}
\begin{figure}
    \centering
    \includegraphics[width=\linewidth]{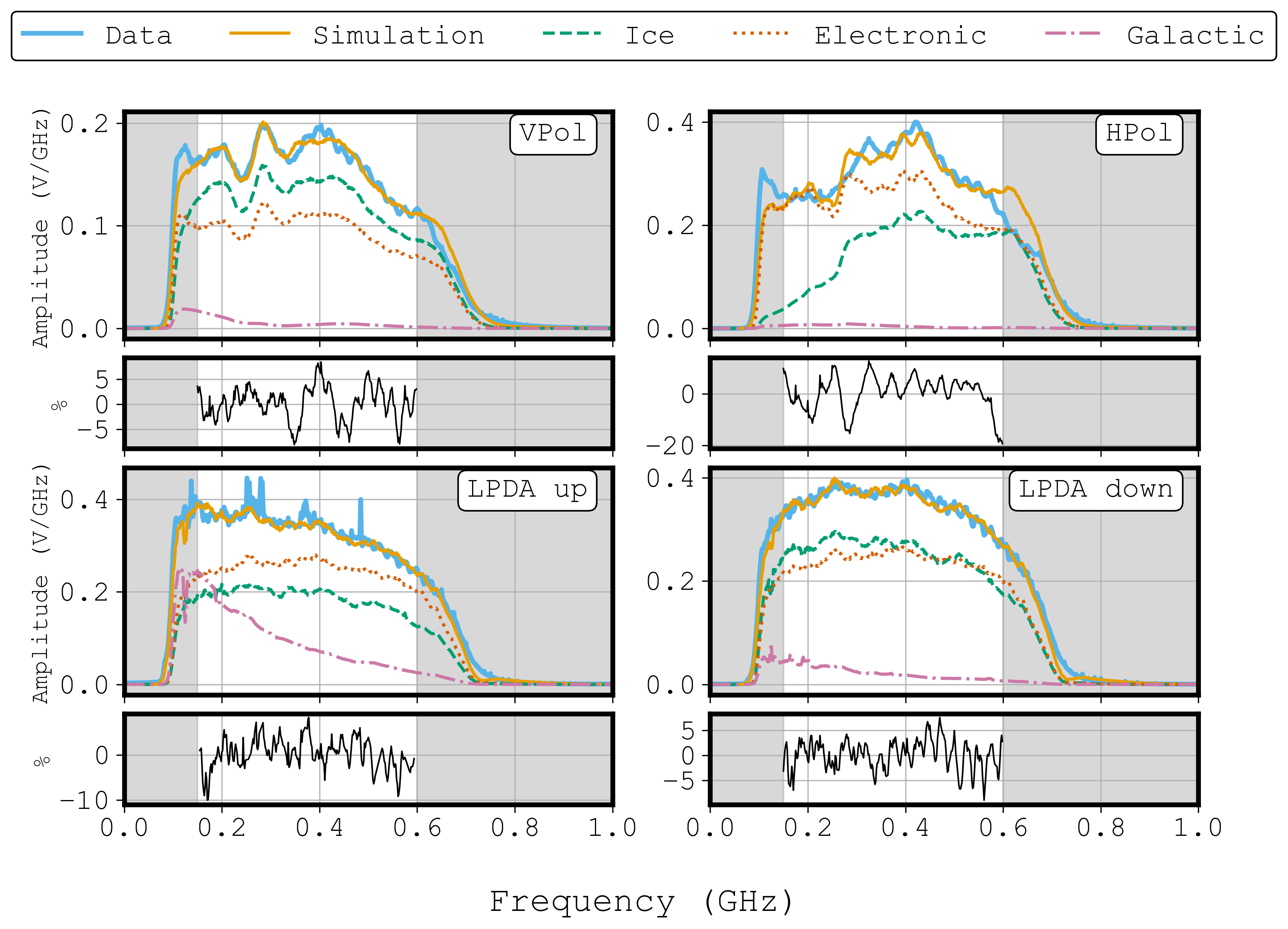}
    \caption{Example fit of our calibration model to data from a full season of RNO-G station 11. Each panel shows, for a different channel, the measured
    and the simulated spectrum (solid lines). The individual signal contributions in our simulation are shown with dashed, dotted, and dash-dotted lines (they add in quadrature, see Eq.~\ref{eq:stot}). The fit range is indicated by the white background. The data--simulation residuals are shown in black, defined as (data $-$ sim)/data. Each channel's spectrum, representative of its type, exhibits its own unique features which our model is able to describe.}
    \label{fig:results}
\end{figure}
The signal model $\Stot$ and the parametric response model $\mathcal{R}(G, s)$ allow us to find an optimal description of the instrumental response of our antenna
channels by fitting the model to the data $\mathcal{D}$ and extracting the absolute gain.
The averaged signal model can be described through
\begin{equation}
    \avg{|\Stot|(f)} = \frac{1}{n} \sum_i^n |\Stot_i(f)| = \frac{1}{n} \sum_i^n | \Sice(f) + \Sele(f) + \Sgal(f) |,
\end{equation}
where $i$ denotes the individual spectra (events). For technical reasons it is better to compare $\avg{|\Stot|^2}$ with data. This has the advantage that the total noise can be described by the averages of the individual components while cross products vanish for uncorrelated noise sources.  

\begin{gather}
    \mathcal{S}'(f) \equiv \sqrt{\avg{|\Stot|^2(f)}} \notag \\[6pt]
    = \Big[ \avg{|\Sice|^2(f)} + \avg{|\Sele|^2(f)} + \avg{|\Sgal|^2(f)}
    + \underbrace{\sum_{k,j;~ k \ne j}^{\text{ice, ele, Gal}} \avg{\mathcal{S}_k(f)\, \mathcal{S}_j^*(f)}}_{\xrightarrow{}\; 0 \,\,\text{(uncorr. noise)}} \Big]^{1/2} \notag \\[6pt]
    \approx \Big[ \avg{|\Sice|^2(f)} + \avg{|\Sele|^2(f)} + \avg{|\Sgal|^2(f)} \Big]^{1/2}.
    \label{eq:stot}
\end{gather}
\noindent
Consequently, to match simulation and data, we derive the mean energy spectral density over the data sample using
\begin{equation}
\label{eq:data}
    \mathcal{D}'(f) = \Big[ \frac{1}{n} \sum_i^n |\mathcal{A}_i(f)|^2 \Big]^{1/2}
\end{equation}
\noindent with $\mathcal{A}_i(f)$ being the real fast Fourier transform (RFFT) for a recorded waveform $A_i(t)$. The frequency-independent gain is optimized to match the recorded data $\mathcal{D}'$ with the simulated data $\mathcal{R} \cdot \mathcal{S}'$:
\begin{equation}
    \mathcal{D}' = \mathcal{R}(G, s) \cdot \mathcal{S}' ~~ \rightarrow ~~ G_\mathrm{opt} = \operatorname*{arg\,min}_{G, s} \big( | \mathcal{D}' - \mathcal{R}(G, s) \cdot \mathcal{S}'|^2 ~/~\text{Var} (\mathcal{D}') \big) 
\end{equation}
The fit is performed with iminuit~\cite{iminuit} as a least-squares minimization within a fit range of 150 -- 600 MHz.
During the fitting procedure we identify narrow-band RFI (i.e., noise peaks in the averaged data spectra) in the upward-facing LPDAs and exclude these band regions from the fit. RFI is identified as peaks in the frequency spectrum that sit above $4$ times the rms, defined as $\sqrt{\sum_{f} \mathcal{D}'(f)^2}$. To avoid a strong bias in the fitted gain due to a correlation with the slope parameter $s$, a three-stage fit is performed\footnote{Such a bias would not be inherently bad as it does not affect the description of the channel response but it would make the comparison of $G$ and $s$ across different channels more difficult.} (cf.~Eq.~\ref{eq:response}). First, $G$ is fitted with $s$ fixed at $0$; then $s$ is fitted with $G$ fixed at the previously fitted value. Finally, both parameters are fitted freely starting from their previously fitted values.

Fig.~\ref{fig:results} shows the fitted model for four channels of station~11, chosen to represent the different antenna-channel types deployed at RNO-G. The fit uses all data of the 2023 season. Solid lines show the data $\mathcal{D}'$ and the model $\mathcal{R}(G, s) \cdot \Stot'$, while the dashed, dotted, and dash-dotted lines show the contributions of the three individual signal components.

The top-left panel shows a phased-array VPol channel. The model describes the data well across the band and reproduces the characteristic dip and peak at approximately 250\,MHz and 300\,MHz. For channels at a depth of about 100\,m, the Galactic contribution to the total signal is negligible, whereas electronic noise and thermal noise from the surrounding ice are comparable over most of the frequency range.

The HPol channels (example top right) exhibit the largest data--simulation discrepancy. The sharp double-peak structure between 300\,MHz and 500\,MHz is present in the model but less pronounced, while the spectrum as a whole is described less accurately compared to the other channel types. We attribute this to uncertainties in the antenna model, which are discussed in Sec.~\ref{sec:antenna_uncert}. In addition, some HPol channels show a strong peak at the lower edge of the band near 100\,MHz that the simulation does not reproduce; unmodeled impedance-matching effects are again a plausible explanation. The total signal in the HPol channels is dominated by electronic noise from the instrument itself, a consequence of the lower sensitivity of this antenna type compared to the VPols and LPDAs.

Both the upward- and downward-facing LPDAs show excellent data--simulation agreement. Apart from weak modulations in the measured spectra that are absent in simulation, and narrow-band RFI peaks in the upward-facing LPDAs, the overall shape and magnitude of the spectra are reproduced accurately. As expected, the Galactic noise contributes more strongly to the upward-facing LPDAs, where it slightly dominates at low frequencies. For the downward-facing LPDAs, thermal ice noise and electronic noise are comparable, as for the VPols, whereas for the upward-facing LPDAs the thermal ice contribution is smaller and electronic noise dominates the total signal.

\begin{figure}
    \centering
    \includegraphics[width=\linewidth]{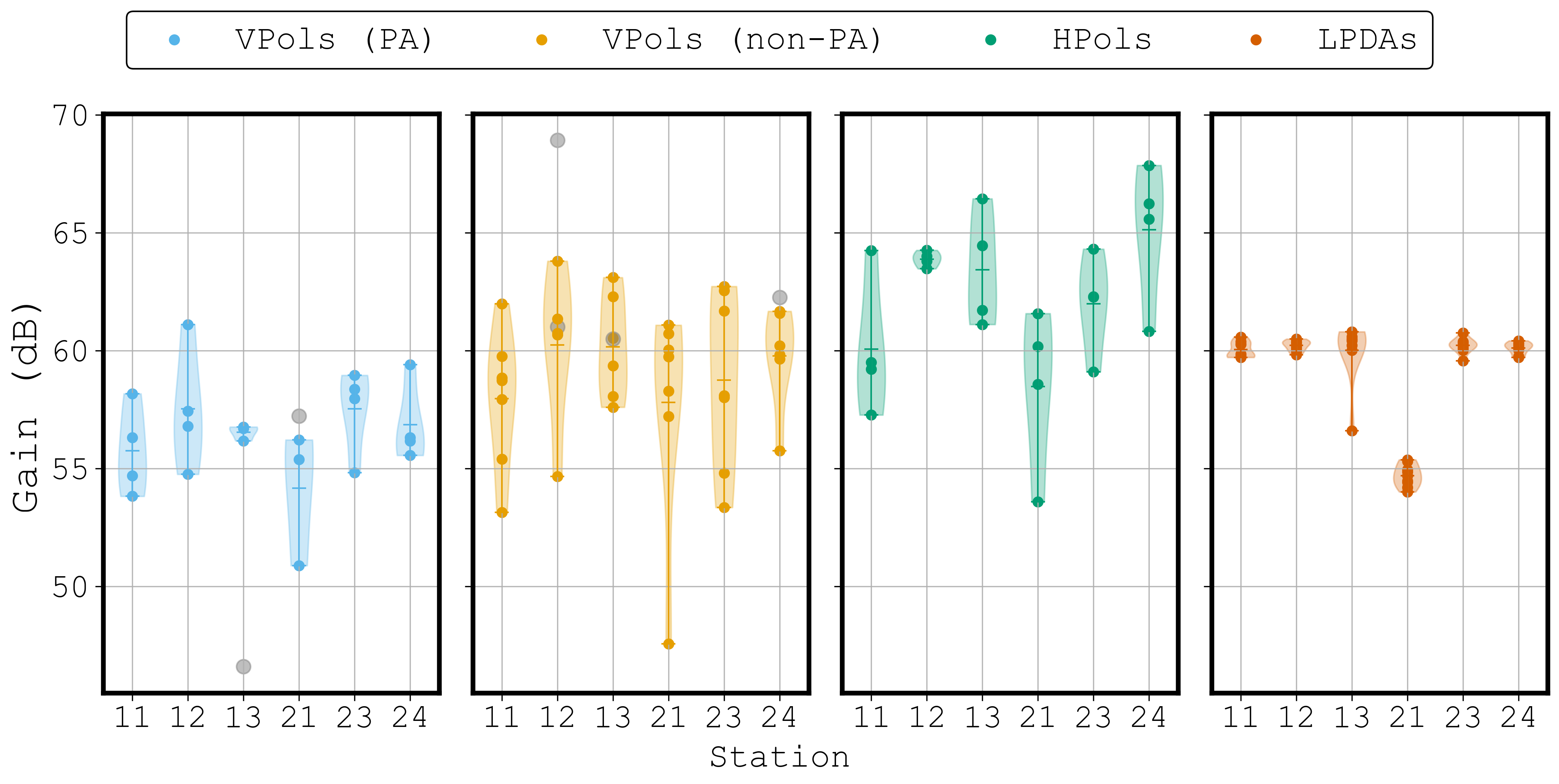}
    \caption{Violin plot of calibrated gains $G$ per station as defined in Equation \ref{eq:response}. The channels are categorized by their respective antenna types.}
    \label{fig:fitted_gain}
\end{figure}

Fig.~\ref{fig:fitted_gain} shows the fitted gains $G$ for all nominally operating channels and stations per channel type in 2023 (broken channels, those with a visibly very different frequency spectrum in data,  are shown with gray markers). In most cases, the fitted gain is close to the design goal of 60\,dB. The deep channels (VPols \& HPols) exhibit a significant scatter of $\sim$10\,dB (or a factor $\sim3$ in amplitude). We attribute this spread to a varying efficiency in the RFoF system: the efficiency of the laser and photodiode as well as signal loss in the fiber connections. The surface channels (LPDAs) show a much narrower distribution close to 60\,dB. A notable outlier is station 21, where all LPDA channels exhibit gains approximately 5 dB lower than in the other stations.

\subsection{Time variability}
\label{sec:time}
\begin{figure}
    \centering
    \includegraphics[width=\linewidth]{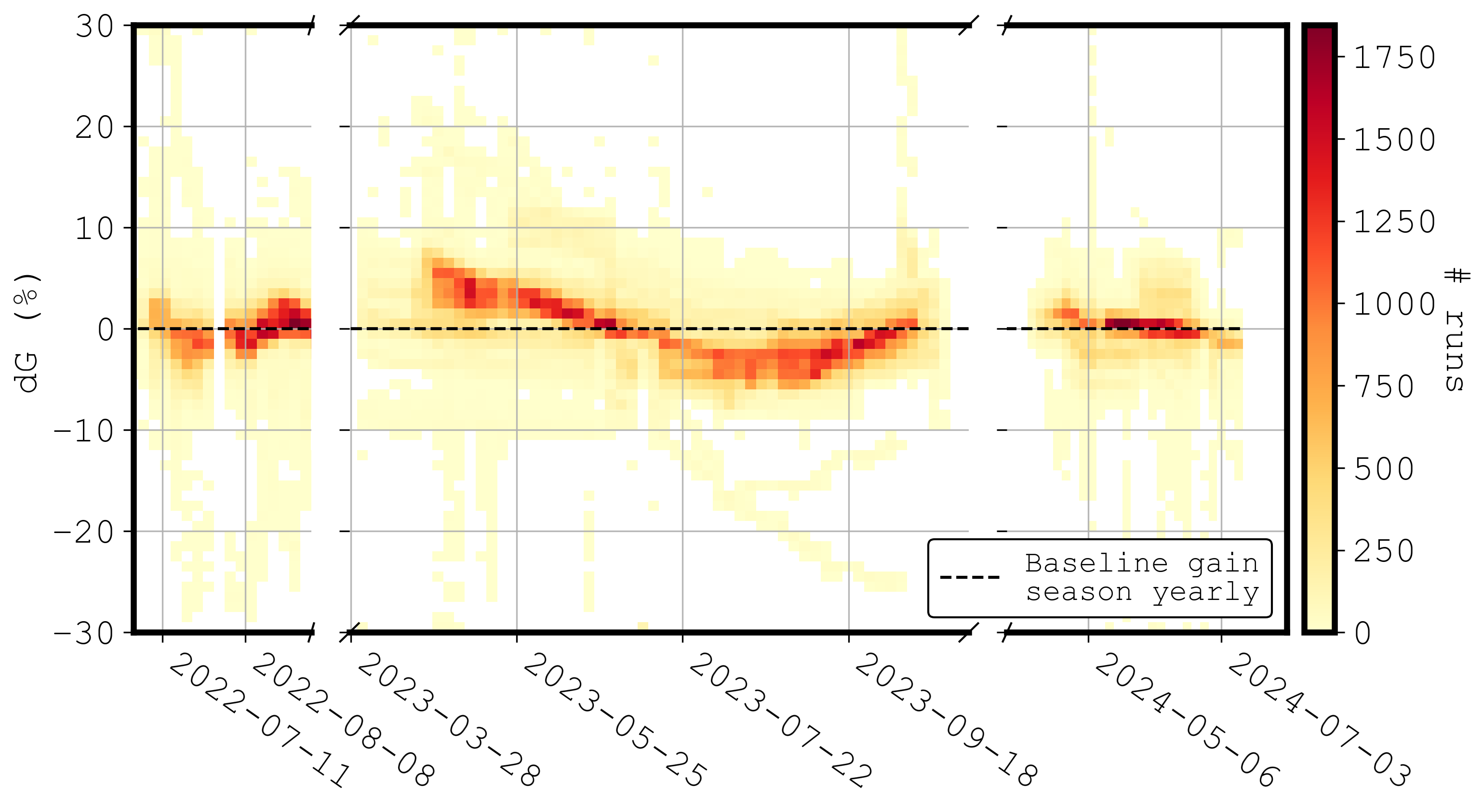}
    \caption{Per-run variation $dG$ (in \%) of the calibrated gain with respect to the nominal seasonal calibration, for all antennas and stations.}
    \label{fig:time_variation}
\end{figure}

The calibration model does not include any time dependence. For the different components in \Stot, any time dependence is either negligible (\Sgal, \Sice) or covered by our uncertainty (\Sele) (see Sec.~\ref{sec:uncertainties}), while the templates are static. 
A time dependence of the channel response could, however,  arise from a temperature-dependent amplifier gain or from aging; the latter is not expected given the short operation time and was not observed by AERA over 7 years of operation~\cite{PierreAuger:2025hoe}.    
To quantify the time variability, we determine the absolute gain per 2-hour run. Fig.~\ref{fig:time_variation} shows the relative deviation $dG = (G_\mathrm{run} - G_\mathrm{season}) / G_\mathrm{season} \cdot 100$ for all channels of six calibrated stations (11, 12, 13, 21, 23, 24). 
We observe three effects:
\begin{itemize}
    \item \textbf{Seasonal modulation.} Within a season, the gain of all nominally operating channels is anti-correlated with temperature, with a rough slope of $-0.25\%$ per \textdegree C. The modulation is most visible in 2023, which covers the longest period.
    \item \textbf{Inter-season shifts.} Between seasons, the gain of individual channels shifts by up to $\sim$20\%. These shifts vary randomly in magnitude and sign from channel to channel, which rules out temperature as their sole cause. For this reason, $dG$ is normalized separately for each season.
    \item \textbf{Unexplained changes.} We also observe sudden gain changes within a season, one of which coincides with a power cycle of the RADIANT board, and a few channels with a gradual drift stronger than expected from temperature alone. Neither effect is correlated with temperature, and their origin is not yet understood. These sudden changes also limit the precision of the temperature slope quoted above.
\end{itemize}
With a per-season averaged gain, which marginalizes over intra-season temperature effects and the unexplained changes, the calibration reaches an accuracy of $\lesssim 5\%$ ($\lesssim 0.5$\,dB). A time-dependent calibration, e.g., per run or per stable period, could reduce this further.

\subsection{Validation}
\label{sec:valid}
To demonstrate the effect of the calibration, we consider the $V_\mathrm{RMS}$ of minimum-bias data, as well as the amplitude of calibration pulser events. 

\subsubsection{\texorpdfstring{$V_\mathrm{RMS}$}{VRMS}}
\label{sec:valid_vrms}

\begin{figure}
    \centering
    \includegraphics[width=\linewidth]{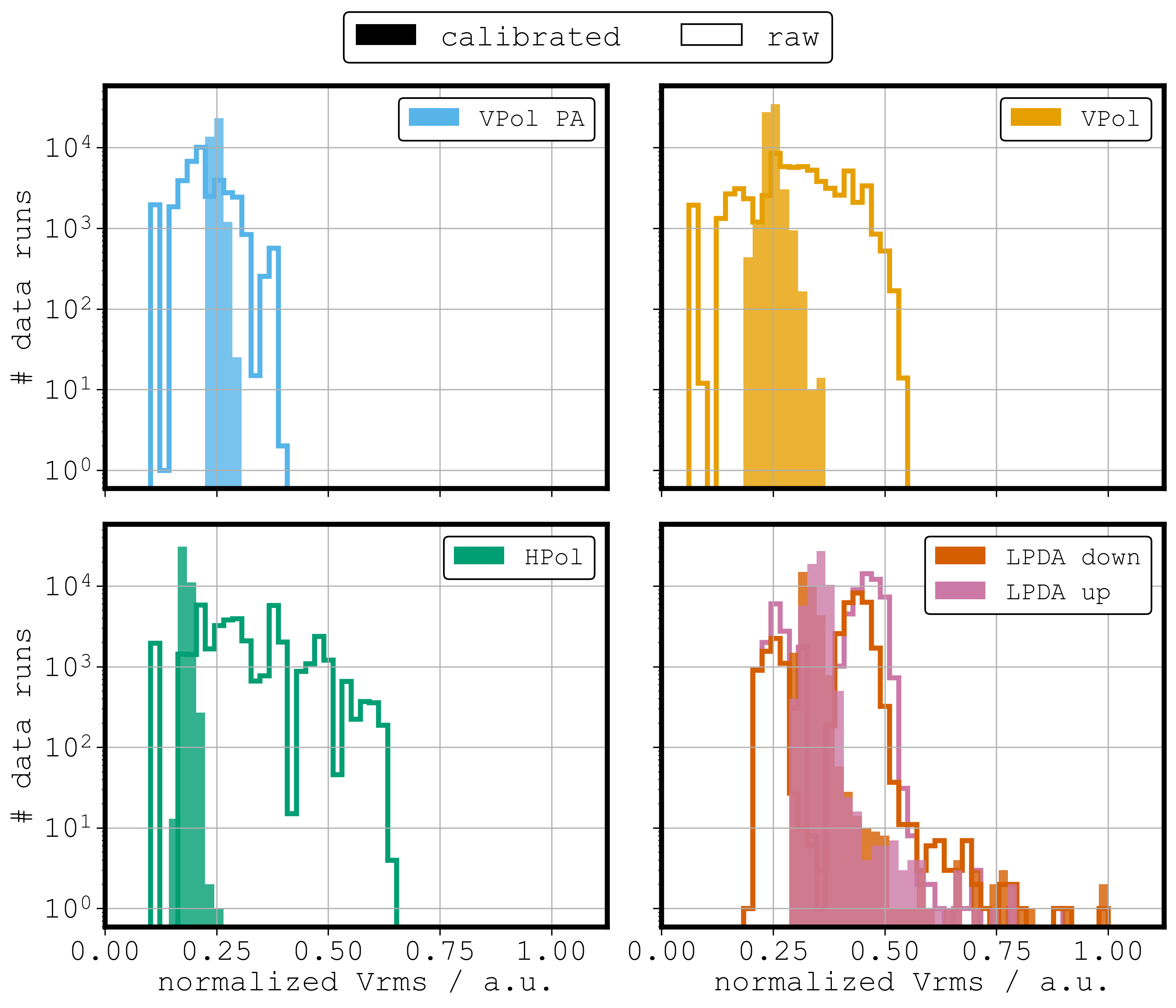}
        \caption{Normalized distribution of data
        $V_\mathrm{rms}$ averaged over 2-hour data-taking periods for all stations over the 2023 season. Solid distributions show $V_\mathrm{rms}$ after dividing by the channel-specific calibrated gain $G$, open distributions the values before calibration. The normalization is chosen such that the maximum value across all panels is $1$.}
    \label{fig:vrms}
\end{figure}

It is instructive to inspect the calibrated $V_\mathrm{RMS}$ and compare it with the uncalibrated case. This can be achieved by normalizing the distribution of $V_\mathrm{RMS}$ values to $\mathrm{max}\left(V_\mathrm{RMS}\right) = 1$ for calibrated and uncalibrated data. In the uncalibrated case, we expect a wide distribution as the distributions of individual channels differ systematically. In the calibrated case, i.e., where the measured $V_\mathrm{RMS}$ are first normalized by the fitted gain, the systematic shift between individual channels (within a group of channels that measure the same physical signals) should disappear and result in a much narrower and well-behaved distribution. This is shown in Fig.~\ref{fig:vrms}.

\subsubsection{Calibration pulsers}
\label{sec:cal_pulser}

To evaluate the effect of the calibrated signal chain on external impulsive signals, we investigate the amplitude of signals emitted by a calibration pulser and received by the phased-array antennas. Given the close proximity of the four antennas, we expect the calibrated signal amplitudes measured by each antenna to be similar.

\begin{figure}
    \centering
    \includegraphics[width=\linewidth]{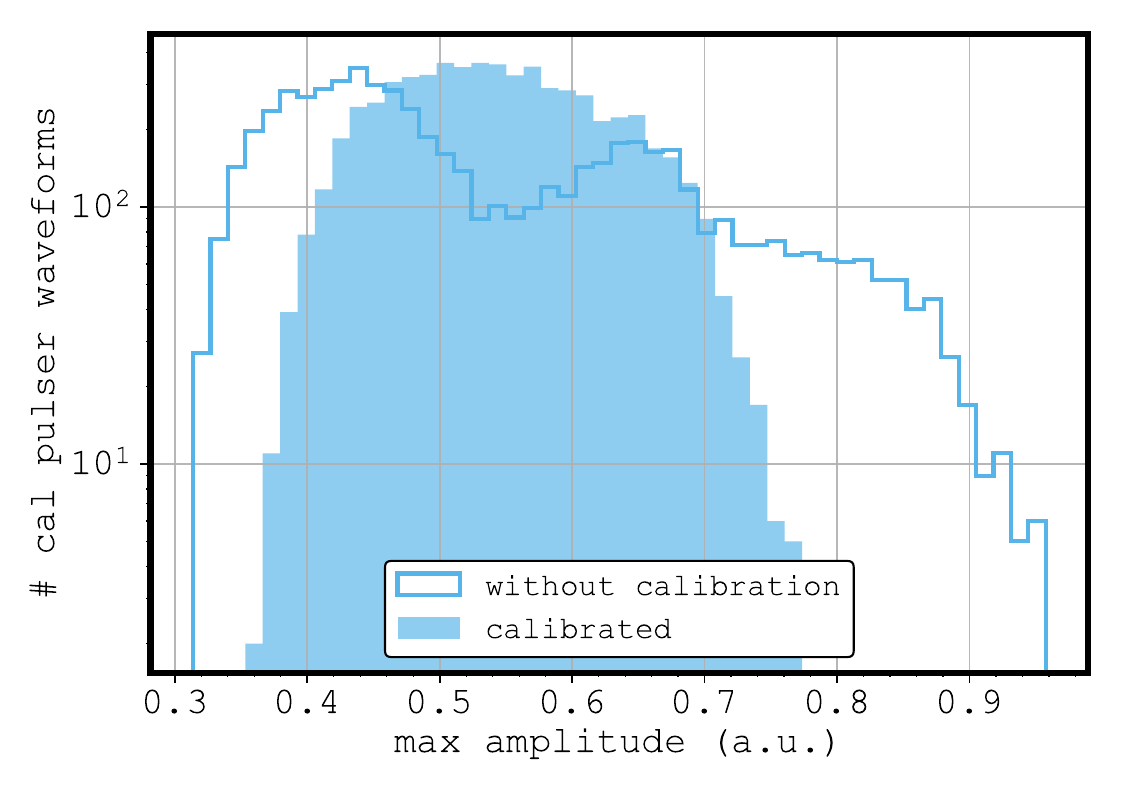}
    \caption{Histogram of the calibration pulser signal amplitudes at the antenna feed for a data-taking run in station 13. The calibrated case (filled) exhibits a single distinct population, while the uncalibrated case (unfilled) exhibits two distinct populations.}
    \label{fig:cal_pulser}
\end{figure}

To verify this expectation, we select a calibration pulser run and unfold the signal-chain response by dividing the measured signal amplitude in each antenna by its corresponding signal-chain gain. This procedure is performed for both the calibrated and uncalibrated detector configurations, with the results shown in Fig.~\ref{fig:cal_pulser}.

We observe a difference in the amplitude distributions between the calibrated and uncalibrated signal chains. While the calibrated detector produces amplitudes clustered around a single value, the uncalibrated detector yields a broader distribution, which in this particular run is even separated into two distinct populations. This behavior indicates a relative gain difference between one of the four antennas and the remaining three. The calibrated signal-chain response accounts for this difference, resulting in consistent reconstructed amplitudes across the four antennas.

\subsection{Impact on energy reconstruction}
\label{sec:energy}

To illustrate the consequences of an inaccurate signal-chain description for physics analyses, we study its impact on neutrino energy reconstruction. Since the true signal-chain response is unknown, this study does not validate the calibration; rather, it quantifies how strongly reconstructed energies depend on the assumed detector description.

We simulated $10^6$ neutrino events with NuRadioMC, with energies distributed uniformly in logarithmic bins between $10^{16}$ and $10^{19}$\,eV. Neutrino interactions are forced within the Greenlandic ice sheet and weighted by their survival probability; the Askaryan emission is derived from template charge-excess profiles and propagated through the ice to the antennas. The full detector response, including the antenna response, the calibrated signal-chain response derived in this work, and the digitization, is applied to obtain simulated waveforms. A trigger threshold of $50$ mV\footnote{In these simulations we assume a flat noise temperature of 300\,K. This translates to a $\sim 5\sigma$ trigger threshold where $\sigma$ is the $V_\mathrm{rms}$ of that noise.} in the triggering antenna was imposed to select events with sufficient signal-to-noise ratio for reliable energy reconstruction.

\begin{figure}
    \centering
    \includegraphics[width=1\linewidth]{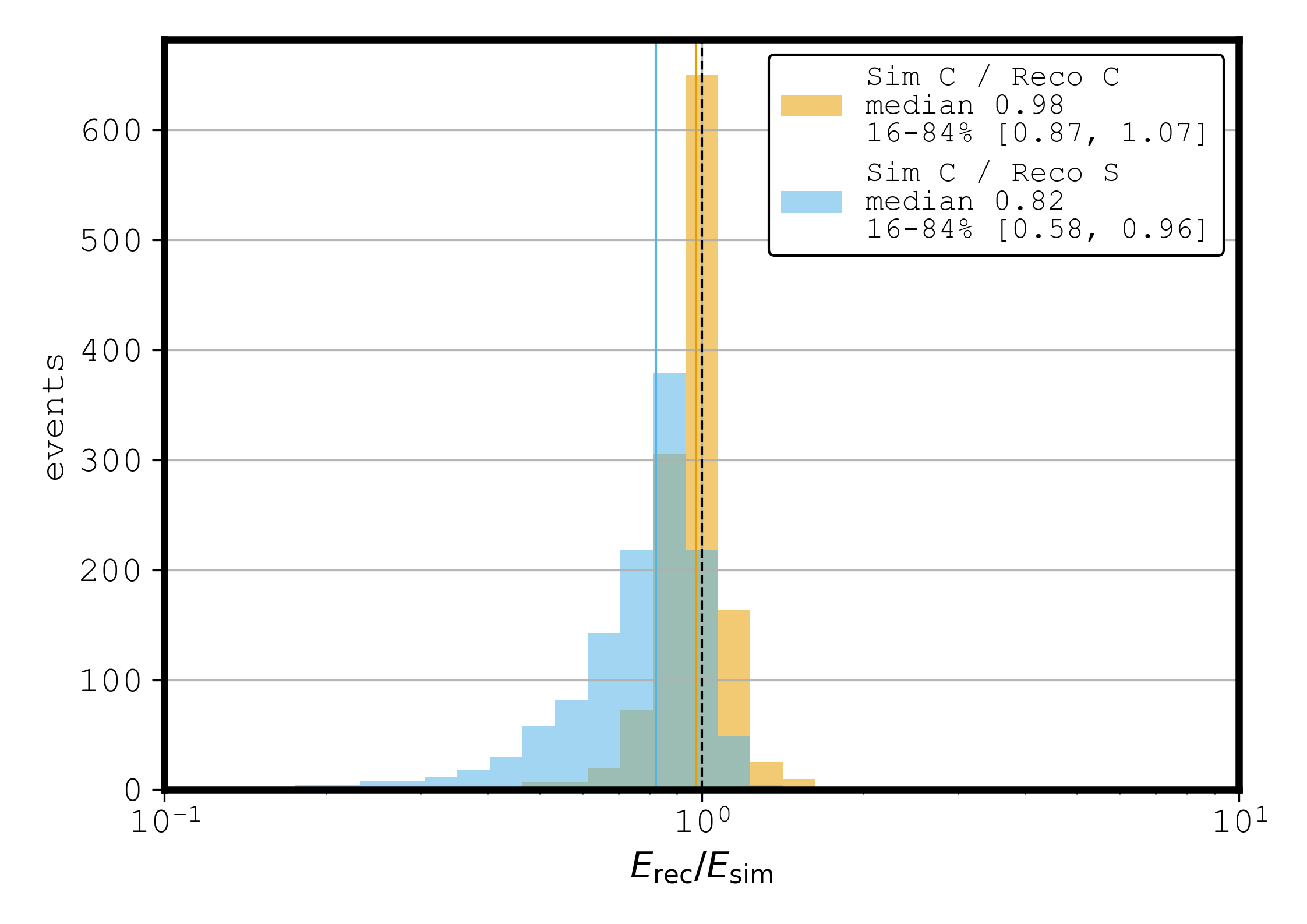}
    \caption{Distribution of the ratio between reconstructed and true simulated neutrino energy. Events are simulated with the calibrated signal-chain response and reconstructed with either the calibrated response (C, orange) or the nominal response obtained by concatenating the $S_{21}$ parameters of the individual components (S, blue).} 
    \label{fig:E_reco}
\end{figure}

The neutrino energy is reconstructed following the methodology described in~\cite{RNO-G_energy}. To isolate the impact of the detector description from uncertainties associated with the reconstruction itself, we use the true neutrino interaction vertex from the simulation rather than reconstructing the vertex from the radio data. Each event is reconstructed twice: once with the calibrated signal-chain response (C), i.e., the response used in the simulation, and once with the nominal, uncalibrated response (S), obtained by concatenating the lab-measured $S_{21}$ parameters of the individual signal-chain components (cf.\ Sec.~\ref{sec:response}). This response differs from the calibrated one in both gain and phase. The results are shown in Fig.~\ref{fig:E_reco}. With the calibrated response, the reconstructed energy is nearly unbiased, with a median ratio $E_\mathrm{rec}/E_\mathrm{sim} = 0.98$ (16--84\% range: 0.87--1.07). With the nominal response, the median shifts to 0.82 and the 16--84\% range widens to 0.56--0.96. A mis-modeled signal chain thus introduces a bias of about 16\% and doubles the spread of the distribution, demonstrating the importance of an in-situ calibration for physics analyses.

\section{Uncertainties}
\label{sec:uncertainties}
In this section, we discuss the uncertainties associated with the calibration. These comprise systematic uncertainties arising from the simulation of the standard-candle noise signal, the modeling of the antenna-channel response, and the calibration method itself. The statistical uncertainties of $G$ from the fit are on the order of $1\%$ or less.

To quantify the calibration uncertainty, it is useful to note that not all effects that change the result of our analysis, i.e., the fitted gain parameter $G$, translate into a one-to-one uncertainty in the modeled channel response. For example, we observed that varying the underlying template response $\mathcal{R}_i$ (cf.~Fig.~\ref{fig:response}) changes the fitted gain by up to 3\%. However, this does not correspond to a 3\% uncertainty in the gain, as the interpretation of $G$ depends on the choice of $\mathcal{R}_i$. A more important example is the antenna model. We find that $G$ is highly degenerate with the magnitude of the realized vector effective length. Consequently, any mis-modeling of the antenna response can be partially absorbed by the fitted gain. In principle, this degeneracy could be reduced by the multi-component noise model, which includes electronic noise independent of the antenna model, provided that the spectral shapes of the individual noise components at the antenna terminals are sufficiently different. However, this degeneracy remains strong, although we have not quantified it. As a result, assigning a quantitative calibration uncertainty is challenging.

Table~\ref{tab:uncert} lists the tested uncertainties related to the signal model and calibration method only. Uncertainties related to the modeling of the instrumental response which are difficult to quantify in terms of an uncertainty on this calibration are discussed in Sec.\ \ref{sec:antenna-channel_uncert}.

\subsection{Signal model}
\label{sec:signal_uncert}
The calibration is based on a model of the ambient noise signal composed of several physical contributions. The following subsections discuss the uncertainties associated with each component of this signal model and their impact on the fitted gain.
\subsubsection{Sky noise}

To conservatively estimate the uncertainty on the sky temperature, we scale its temperature down by 5\% and up by 15\% covering the range of predictions at the location and frequency band of RNO-G~\cite{Busken:2022mub}. A 15\% shift in temperature results in a 7.5\% shift in received power. For most antenna channels the emission from the sky is subdominant. Hence, the observed variation in fitted gain with shifted simulations is below $1\,\%$ for all antenna models. This is also true for the upward-facing LPDAs.

\subsubsection{Electronic noise temperature}

The nominal \Sele is based on lab measurements with the downhole amplifier at \tc{-40} and surface amplifier at \tc{0}. While the temperature in the borehole below a depth of 20\,m is expected to be constant\footnote{The temperature in the borehole, based on ice-core measurements, is expected to be approximately \tc{-32}, which is higher than the assumed temperature of \tc{-40}. We consider this difference to be negligible.}, the temperature in the DAQ box can vary during operation and is affected by seasonal variation and snow accumulation. As a conservative estimate of this effect, we repeat the calibration using electronic noise measured at \tc{+20}. A temperature of \tc{+20} (i.e., a shift of 20\textdegree C) is expected to represent the upper end of the temperature range encountered during operation. This results in an uncertainty in the calibrated gain of approximately 2\% for the surface signal chain.

\subsubsection{Ice noise} 
We assessed the robustness of the apparent ice temperature \Teff against the input parameters of the ice model and against the approximations made in Eq.~\ref{eq:teff}.

The ice temperature is tabulated as a function of depth ($z = 1.1, 11.1, 21.1, \ldots$\,m) and interpolated in between. To estimate the impact of the seasonal temperature variations near the surface, we varied the values of the top two layers, i.e., the temperature in the upper 20\,m, between \tc{-40} and \tc{-10} around the nominal value of \tc{-31.7}. The effect is negligible: owing to the kilometer-scale attenuation length, \Teff is an integral over a large ice volume and is dominated by the average temperature of the bulk ice, which is constant and well known.

Since $l_\mathrm{att}$, and consequently the emission coefficient, depends on frequency, \Teff is, strictly speaking, also frequency dependent. The attenuation length determines how deep into the ice the antenna is sensitive and thus modulates \Teff as a function of the signal arrival direction. We find that this changes \Teff by at most 2\% for the shallow LPDAs (less for the deep antennas), which corresponds to less than 1\% in electric-field amplitude; the effect on the fitted gain is smaller still. We therefore evaluate \Teff at a fixed frequency of 400\,MHz.

Variations in the antenna deployment depth, particularly for the surface antennas, have no noticeable impact, and neither does the index-of-refraction model used for the ray tracing. Given the weak sensitivity to all of the above, we expect other potential sources of uncertainty, such as the choice of attenuation-length model, to be similarly negligible.

Equation~\ref{eq:teff} neglects scattering and focusing. Focusing is strongest for near-horizontal trajectories close to the shadow zone, where the ice temperature is nearly uniform. Since an extended source of homogeneous temperature is not enhanced by focusing, we consider this effect negligible.

The reflection coefficient $|r_i|^2$ is calculated by summing over all polarizations. While the thermal emission from the ice is unpolarized, small differences between the polarization-dependent reflection coefficients would slightly polarize the reflected signal. We neglect this effect, as it is small and the reflected emission contributes little to the total signal at the antenna.

\begin{table}[t]
    \centering
    \caption{List of tested uncertainties with respect to the signal model and calibration method.}
    \vspace{0.5cm}

    \noindent\makebox[1\textwidth]{%

        \begin{tblr}{
          colspec = {Q[c,m,wd=3.2cm] Q[c,m,wd=1.5cm] X[c,m]},
          hline{2-Y} = {solid},   % inner horizontal lines only
          vline{2-Y} = {1pt, solid},   % inner vertical lines only
          hline{2}   = {1pt, solid},    % thick line under the header
          width   = 1\textwidth,           % <-- wider than the text block
          rowsep  = 3pt,
          % hlines  = {black},
          % vlines  = {black},
          column{3} = {font=\footnotesize},  
          % row{1}    = {font=\bfseries},   % header
          %column{1} = {font=\bfseries},   % Source column bold
          % cell{2-Z}{1} = {font=\small},   % overrides column{1} font from row 2 down
        }
            Source & Value & Remark \\
            Fit range & $<1\%$ & Fit range varied within 100--700\,MHz \\
            \Sgal & $\ll 1\%$ & --- \\
            Refractive index $n$ & $<1\%$ & Using \rVEL simulations at different $n$; $\Delta G \approx  \frac{\Delta n}{0.1}\%$
            \\
            Temperature-dependent gain  & $<5\%$ & Can be reduced by a time-dependent gain description. \\
            \Sice & $<1\%$ & Seasonal fluctuations in the first 20\,m; frequency dependence of attenuation length. \\
            \Sele & 2\% & Temperature dependence of the noise (not the gain). Only for LPDAs. \\
        \end{tblr}
    }
    \label{tab:uncert}
\end{table}

\subsection{Antenna-channel response}
\label{sec:antenna-channel_uncert} 

In addition to the signal model, the calibration depends on the modeling of the antenna-channel response. The following subsections discuss the dominant uncertainties associated with the antenna model and the impedance mismatch.
\subsubsection{Antenna models}
\label{sec:antenna_uncert}

The VPol and HPol antennas are modeled using the finite-difference time-domain (FDTD) solver XFdtd \cite{xfdtd}. For this analysis, we adopt version~4 of the VPol model and version~5 of the HPol model, based on the models developed in \cite{smith_2022_kzvrc-02b40, RNO-G:2023xyg}. The LPDA model is taken from \cite{Barwick:2014boa} and validated with data. The realized vector effective length (\rVEL) depends not only on the antenna design (i.e., its mechanical structure and materials) but also on the surrounding environment, including the borehole geometry (and, consequently, the antenna placement within the borehole) and the properties of the surrounding ice. This complexity makes it difficult to validate the simulated \rVEL with, e.g., lab measurements in an anechoic chamber. For the VPols and HPols, the antenna response was simulated both in air and in ice (i.e., for an ice-surrounded borehole). The former can be directly compared with measurements performed in an anechoic chamber. A comparison with measurements performed by the collaboration is presented in Appendix~\ref{app:antenna}. Overall, the simulated \rVEL and voltage standing wave ratio (VSWR) show good agreement with data. While the magnitude of the \rVEL is reproduced well, the comparison suggests a possible shift in the resonance frequency of the VPol model. To investigate this effect, we shifted the entire VPol model in frequency by $\pm 50\,$MHz and repeated the analysis. We found the resulting fitted gain $G$ to vary by $|\Delta G| \lesssim 8\%$. 

One ice property that affects the antenna response, both in terms of its overall sensitivity and its resonance frequency, is the refractive index of the surrounding ice. Using simulations with refractive indices ranging from 1.0 to 1.8 in steps of 0.1, we established for the VPols that the fitted gain changes by approximately 1\% per $\Delta n = 0.1$. Since the VPol antennas are deployed at different depths, antenna patterns corresponding to the appropriate local refractive index are used for each antenna. The residual uncertainty in the refractive index is significantly smaller than 0.1 and therefore results in a negligible uncertainty in the antenna response. 

Anechoic chamber data for the HPol are limited but show reasonable agreement within the relevant frequency range. Since the HPol antenna is less sensitive than the other antenna models, the signal-chain gain fitted in this analysis is expected to be better constrained, i.e., to exhibit a smaller degeneracy with the antenna sensitivity. However, the agreement between simulation and data is poorest for the HPol antennas. This may be partially due to the residual discrepancies between 300\,MHz and 400\, MHz visible in Fig.~\ref{fig:rvel} and partly due to the fact that impedance matching is more challenging for the HPol antennas than for the VPol antennas, as shown in Appendix~\ref{app:impedance}.

\subsubsection{Impedance matching}

In our modeling of the full instrumental response, we model the impedance matching of the simulated antenna models to a \SI{50}{\Omega} receiver (LNA). Any additional mismatch arising from the fact that the receivers are not perfectly matched to \SI{50}{\Omega} is not modeled. In Appendix~\ref{app:impedance}, we derive a correction factor and calculate it for the VPol and HPol models and one IGLU load impedance measurement. Applying this correction to a representative dataset resulted in a negligible change in the fitted gain. This is because the correction is significant only at low frequencies, which provide limited sensitivity to the frequency-independent gain parameter. Nevertheless, accurate modeling of the impedance mismatch could improve the agreement between data and simulation in the future, particularly at low frequencies.

\section{Conclusion}
\label{sec:conclusion}

This work presented an in-situ calibration of the RNO-G signal-chain response based on measurements of the ambient detector noise. We demonstrated that comparing first-principles simulations of the physical noise contributions with a parametric model of the detector response yields a calibrated description of the instrument that can be used in subsequent collaboration efforts, including detector simulations, event reconstruction, and physics analyses. This is the first application of a multi-component noise calibration, originally developed for radio air-shower experiments, to an in-ice radio neutrino detector, providing a blueprint for the application of this calibration method to future in-ice radio detectors such as IceCube-Gen2~\cite{IceCube-Gen2:2020qha, Gen2}.

The fitted gains of the surface channels cluster closely around the design value of 60\,dB, whereas the deep channels show a spread of about 10\,dB, which we attribute to variations in the efficiency of the RF-over-fiber links. Per-run fits reveal an anti-correlation of the gain with temperature of about $-0.25\%$ per \textdegree C, as well as shifts of up to $\sim$20\% between seasons. With a season-averaged gain per channel, the calibration reaches an accuracy of $\lesssim5\%$ ($\lesssim0.5$\,dB); a time-dependent calibration could improve this further.

Systematic uncertainties from the signal model are small: variations of the sky model, the ice model, and the electronic-noise temperature change the fitted gain by at most 2\%. The dominant limitation is the antenna model, since the realized VEL is largely degenerate with the fitted gain. The variations of the antenna model considered here changed the fitted gain by up to 8\%; because of this degeneracy, however, they cannot be directly interpreted as an uncertainty on the calibrated gain.

The calibration framework presented here establishes a robust and data-driven method for monitoring and characterizing the response of the RNO-G detector throughout its operation. The method can be extended to any hardware revision for which a full-chain lab measurement of the system response is available. This is already the case for the digitizer board of station~14, while dedicated measurements of the new DAQ deployed in the 2026 field season are still required. As additional calibration data and improved detector models become available, the approach can be refined further by incorporating more accurate descriptions of the antenna response, impedance matching, and environmental conditions. Beyond improving the detector description, the method provides a systematic framework for validating detector simulations against measured ambient noise and for tracking changes in the instrumental response over time.

\acknowledgments

We are thankful to the support staff at Summit Station for making RNO-G possible. We also acknowledge our colleagues from the British Antarctic Survey for having built and operated the BigRAID drill for our project.

We would like to acknowledge our home institutions and funding agencies for supporting the RNO-G work; in particular the Belgian Funds for Scientific Research (FRS-FNRS and FWO) and the FWO programme for International Research Infrastructure (IRI), the National Science Foundation (NSF Award IDs 2411590, 2514262, 2514201, 2514208, 2514206, and collaborative awards 2310122 through 2310129), and the IceCube EPSCoR Initiative (Award ID 2019597), the Helmholtz Association, the German Research Foundation (DFG, Grant 556092823), the Swedish Research Council (VR, Grant 2023-00156), the University of Chicago Research Computing Center, and the European Union under the European Union's Horizon 2020 research and innovation programme (ERC, Pro-RNO-G No 101115122 and NuRadioOpt No 101116890). 

Ruben Camphyn is a FRIA grantee of the Fonds de la Recherche Scientifique – FNRS

\bibliographystyle{JHEP}
\bibliography{biblio}

\appendix

\section{Code}
\label{app:code}
This analysis is mostly written in Python and uses several popular analysis libraries (scipy, iminuit, numpy, matplotlib). The simulation of the background radiation and detector response was performed with NuRadioReco (a sub-package of the NuRadioMC framework). The following list includes specific modules in NuRadioMC which were developed, improved, and/or used in this analysis:
\begin{itemize}
    \item \href{https://github.com/nu-radio/NuRadioMC/blob/develop/NuRadioReco/modules/channelIceThermalNoiseAdder.py}{channelIceThermalNoiseAdder}: Simulates \Sice for a given temperature profile (see next bullet). Developed in this work.
    \item \href{https://github.com/nu-radio/NuRadioMC/tree/develop/NuRadioMC/examples/simulate_effective_ice_temperature}{simulate\_effective\_ice\_temperature}: Uses \texttt{RadioPropa} to calculate the effective ice temperature. Developed in this work.
    \item \href{https://github.com/nu-radio/NuRadioMC/blob/develop/NuRadioReco/modules/channelGalacticNoiseAdder.py}{channelGalacticNoiseAdder}: Simulates \Sgal for a specific detector and emission model. Improved in this work.
    \item \href{https://github.com/nu-radio/NuRadioMC/blob/develop/NuRadioReco/modules/channelGenericNoiseAdder.py}{channelGenericNoiseAdder}: Simulates \Sele for a given temperature.
\end{itemize}

\section{Impedance matching}
\label{app:impedance}

In this section we motivate and derive a possible impedance-mismatch correction factor for the antenna-channel response. This correction is not applied in the presented calibration but could improve the description of the full antenna-channel response in the future.

This additional correction factor is needed because the realized vector effective length \rVEL describes the signal from the antenna at a 50\,$\Omega$-matched receiver. However, in reality the IGLU and SURFACE boards are not perfectly 50\,$\Omega$-matched. Hence, we derived a correction factor based on a lab-measured impedance for an IGLU and SURFACE board and simulated impedance of the antennas.

We derived the correction only for one IGLU; in the future, a more faithful approach is to derive channel-specific correction factors (with the shortcoming that we can only use a simulated antenna impedance). Furthermore, the resulting correction factors were mostly flat and only show significant deviations from one outside of our calibration band (cf.\ Fig.\ \ref{figure:antenna-amplifier-mismatch-s11-cable}). Hence, this correction is omitted here but a future calibration / description of the instrumental response could account for it by modeling the impedance mismatch.

Some literature does not explicitly make the differentiation between the vector effective length \VEL and realized vector effective length \rVEL (omitting vector notation). While $\mathcal{H}$ describes the open-circuit voltage an incoming electric field produces at the antenna terminal $V_\mathrm{oc} = \mathcal{H} E$, $\mathcal{H}_r$ describes the voltage at the input of a connected amplifier $V_\mathrm{in}$ with an additional term $\rho$:
\begin{equation}
\label{eq:hr}
    V_\mathrm{in} = \underbrace{\rho \mathcal{H}}_{\mathcal{H}_r} \, E ~~~ \mathrm{with} ~~~ \rho = \frac{Z_L}{Z_L + Z_A}
\end{equation}
where $Z_A$ is the antenna impedance, and $Z_L$ is the load (amplifier) impedance. The factor $\rho$ describes losses due to (im)perfect impedance matching between antenna and load. Note that even for perfect matching, $Z_A = Z_L$, only half of the open-circuit voltage reaches the amplifier, i.e., $\rho = 1/2$. It is standard procedure, when modeling an antenna to match it to a reference impedance $Z_L = Z_0 \equiv 50\,\Omega$ ($Z_A$ is calculated from the same code used to calculate $\mathcal{H}$). Hence, the antenna pattern used here translates the antenna signals to the input terminal of a 50\,$\Omega$ receiver $\mathcal{H}_r$ = $\mathcal{H}_r(Z_L=Z_0)$. 

\begin{figure}
	\centering
	\includegraphics[width=\textwidth]{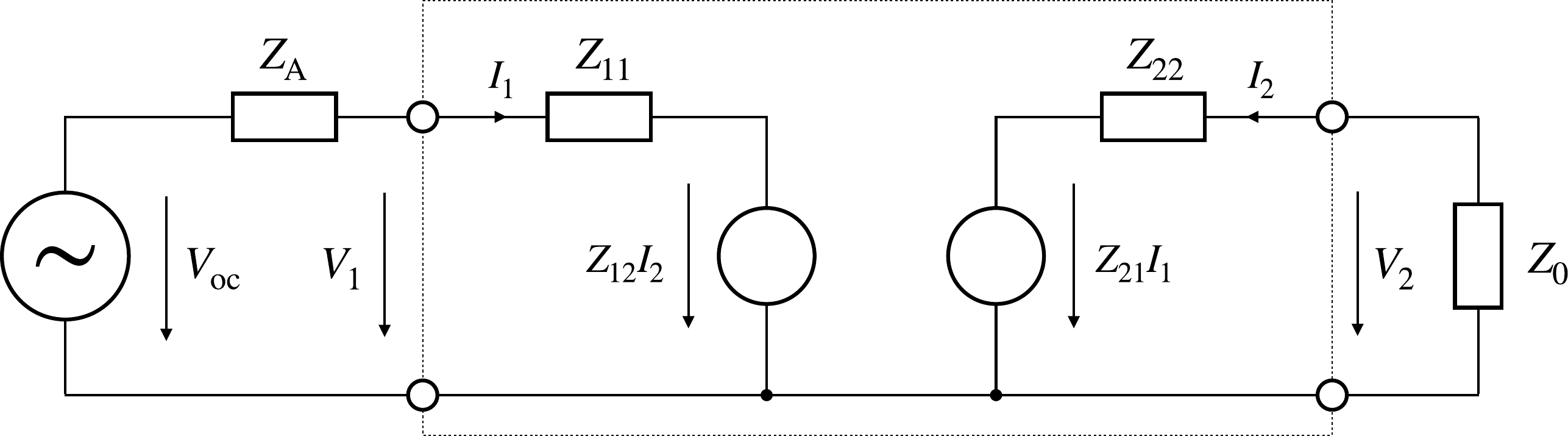}
	\caption{Situation considered here, with an antenna coupled to the input port of the analog signal chain.}
	\label{fig:antenna-coupling-z-pars}
\end{figure}

However, in reality, a receiver (i.e., a low-noise amplifier) does not have an impedance of exactly $50\,\Omega$ (equivalent to $S_{11}^\mathrm{LNA} \neq 0)$. Hence, an additional term to match the simulated antenna to a realistic amplifier is needed. To derive this term, we consider the equivalent circuit of an antenna (parameterized by the open-circuit voltage $V_\mathrm{oc}$ and the antenna impedance $Z_A$) connected to the readout chain (here parameterized in terms of impedance parameters $Z_{11}$, $Z_{12}$, $Z_{21}$, and $Z_{22}$), as shown in Fig.~\ref{fig:antenna-coupling-z-pars}.
We imagine the output port of the signal chain to be connected to a matched load $Z_0$. The impedance parameters relate the voltages and currents as

\begin{equation}
	\left(\begin{matrix}
		V_1\\
		V_2
	\end{matrix}\right)
	=
	\left(\begin{matrix}
		Z_{11} & Z_{12}\\
		Z_{21} & Z_{22}
	\end{matrix}\right)
	\left(\begin{matrix}
		I_1\\
		I_2
	\end{matrix}\right).
	\label{eq:z-par-def}
\end{equation}

The impedance parameters are related to the scattering parameter by:

\begin{align*}
    Z_{11} &= Z_0 \frac{(1 + S_{11}) (1 - S_{22}) + S_{12}S_{21}}{\Delta_S} \\
    Z_{22} &= Z_0 \frac{(1 - S_{11}) (1 + S_{22}) + S_{12}S_{21}}{\Delta_S} \\
    Z_{12} &= Z_0 \frac{2S_{12}}{\Delta_S} \\
    Z_{21} &= Z_0 \frac{2S_{21}}{\Delta_S},
\end{align*}
\noindent where $\Delta_S = (1 - S_{11})(1 - S_{22}) - S_{12}S_{21}$

Solving the Equation \ref{eq:z-par-def}  for $V_2=V_\mathrm{out}$ describing the voltage at the output of the amplifier, we get

\begin{equation}
		V_\mathrm{out} = V_\mathrm{oc} S_{21} \cdot \frac{Z_0}{Z_0 (1+S_{11}) + Z_A (1 - S_{11})}.
\end{equation}

With $V_\mathrm{out} = S_{21} V_\mathrm{in}$ and Eq.~\ref{eq:hr} (substituting $Z_L=Z_0$) we get the required correction factor:

\begin{equation}
\frac{V_\mathrm{out}} {V_\mathrm{out} (S_{11}^\mathrm{LNA} = 0)} = \frac{Z_A + Z_0}{Z_A (1 - S_{11}^{\mathrm{LNA}}) + Z_0 (1 + S_{11}^{\mathrm{LNA}})}.
\end{equation}
However, we can write the original equation \ref{eq:hr} with $\rho$ and the additional correction factor by substituting $Z_A$ with $S_{11}^\mathrm{A}$

\begin{equation}
	\label{eq:antenna-amplifier-mismatch-s11}
	V_\mathrm{out} = S_{21}^\mathrm{LNA} \cdot 
		\underbrace{\frac{1-S_{11}^\mathrm{A}}{2}}_{\rho(Z_L=Z_0)}
	\cdot 
		\underbrace{\frac{1}{ 1- S_{11}^\mathrm{A}S_{11}^\mathrm{LNA}}}_{\mathrm{add.~ antenna-amplifier~mismatch}}
	\cdot 
		\vec{H}(\theta, \phi) \cdot \vec{E}.
\end{equation}

Furthermore, antennas and amplifier are typically connected via a coaxial cable. Because neither the amplifier nor the antennas are perfectly matched to $50 \Omega$, there will be reflections in the coaxial cable, and the cable itself needs to be taken into account. If we assume the cable is lossless and perfectly matched to $Z_0 = 50 \Omega$, its only effect is to introduce a cable delay, i.e.\
\begin{equation}
	t_\mathrm{cable} =  e^{-i\beta L},
\end{equation}
where $\beta = 2 \pi f / u$ is a function of the frequency and the propagation speed $u$ inside the cable, and $L$ is the length of the cable. Then the correction factor is 

\begin{equation}
	\label{eq:antenna-amplifier-mismatch-s11-cable}
	V_\mathrm{out} = S_{21}^\mathrm{LNA} \cdot 
		\underbrace{\frac{1-S_{11}^\mathrm{A}}{2}}_{\rho(Z_L=Z_0)}
	\cdot 
		\underbrace{\frac{e^{-i\beta L}}{ 1- S_{11}^\mathrm{A}S_{11}^\mathrm{LNA} ~ e^{-i\beta L}}}_{\mathrm{add.~ antenna-amplifier~mismatch}}
	\cdot 
		\vec{H}(\theta, \phi) \cdot \vec{E}.
\end{equation}

\begin{figure}
	\centering
	\includegraphics{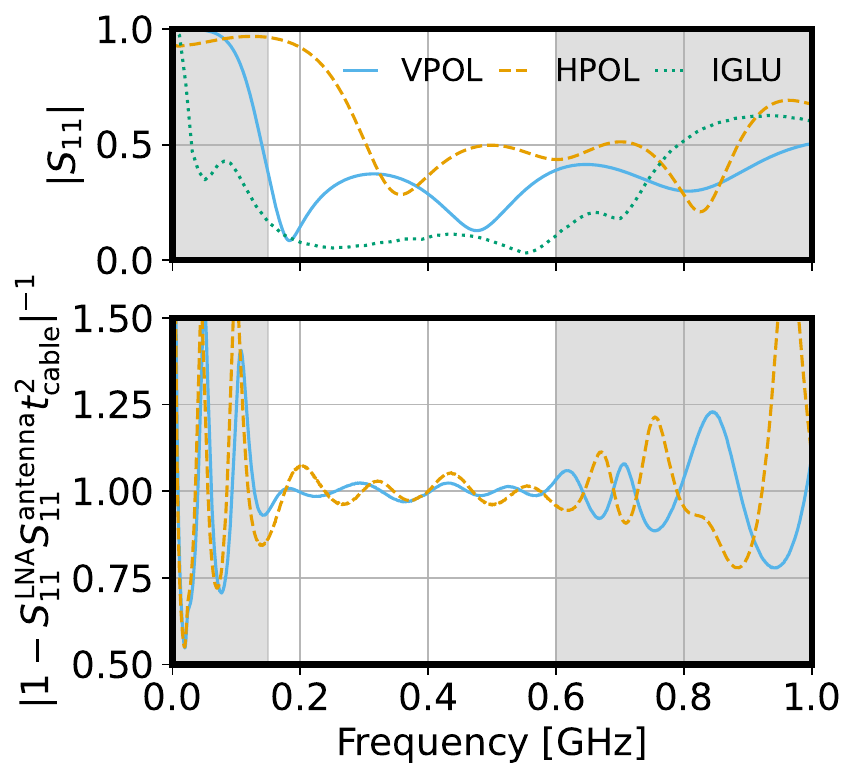}
	\caption{Top: the $S_{11}$ values for the VPol and HPol antennas, as well as the IGLU amplifier. Bottom: correction factors 'antenna-amplifier mismatch' for the VPol and HPol antennas, as defined in Eq.~\ref{eq:antenna-amplifier-mismatch-s11-cable}.}
	\label{figure:antenna-amplifier-mismatch-s11-cable}
\end{figure}

An example for the derived correction factors for the simulated VPol and HPol calculated with a measured LNA impedance are shown in Fig.~\ref{figure:antenna-amplifier-mismatch-s11-cable}.

\section{Antenna model validation}
\label{app:antenna}
To validate the simulated antenna models, in-air simulations can be compared to anechoic chamber measurements. The utilized models were compared to data from the anechoic chamber at the University of Kansas. Details about this measurement campaign are given in \cite{RNO-G:2023xyg}.

\begin{figure}
	\centering
	\includegraphics[width=0.7\textwidth]{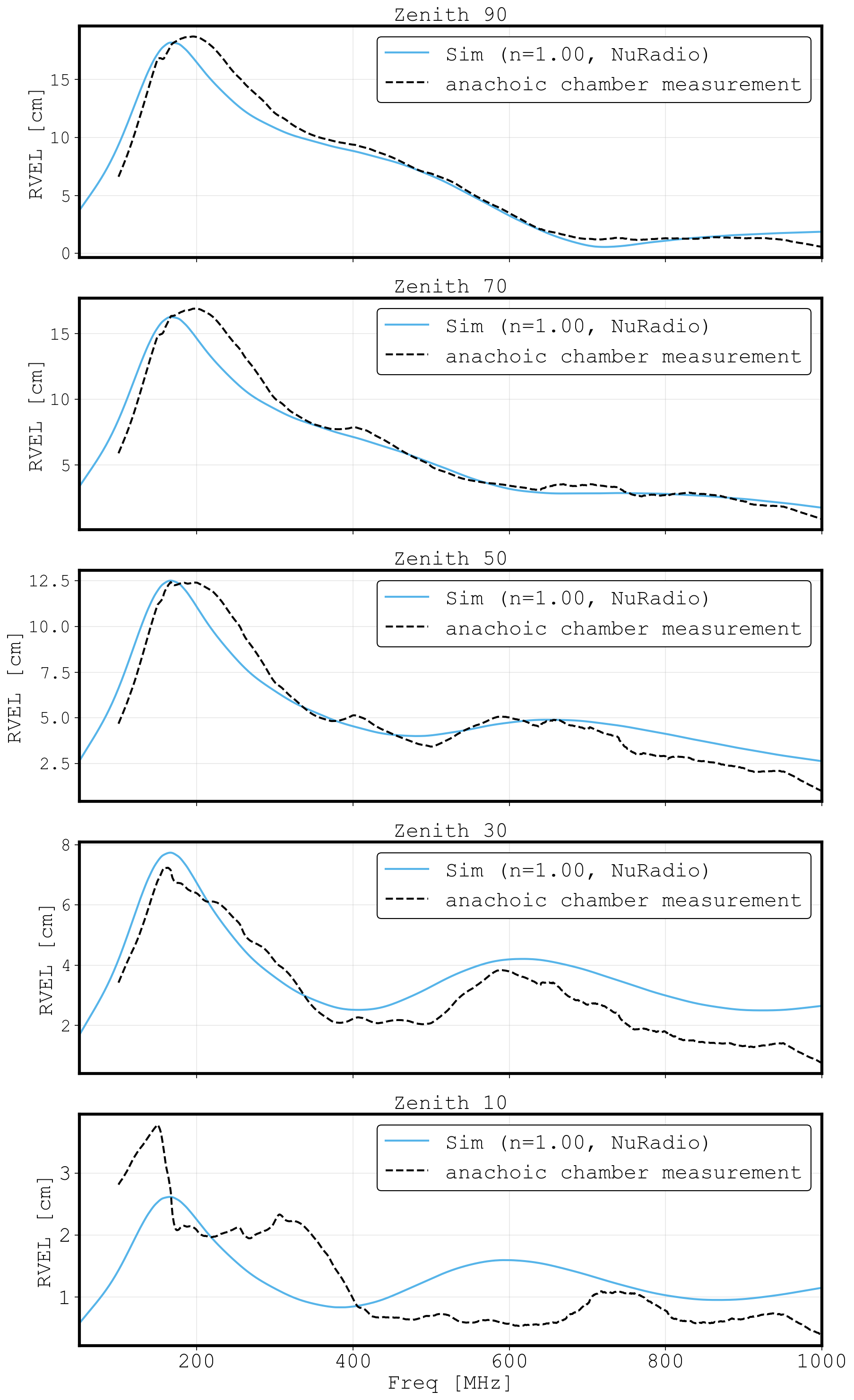}
	\caption{Comparison of simulations of the realized vector effective length \rVEL for VPol antennas with anechoic chamber measurements. Good agreement is found for most signal directions in the RNO-G relevant bandwidth of approx.\ 100\,MHz to 600\,MHz. Shown are VPol at different angles from bore-sight.}
	\label{fig:rvel}
\end{figure}

The following plots show the comparison between the latest iteration of our VPol (left) and HPol (right) models with data. Fig.\ \ref{fig:rvel} shows the magnitude of the realized vector effective length as a function of the frequency for different off-axis angle to the bore-sight (VPol, see panel titles) and at bore-sight (HPol). Fig.\ \ref{fig:vswr} shows the voltage standing wave ratio for both antenna types as a function of the frequency. It should be noted that the impedance matching networks are designed to give the best match when the antennas are in the borehole at their target depth. An in-air anechoic chamber measurement therefore does not actually probe the situation that is relevant for detector operation.

\begin{figure}
    \centering
	\includegraphics[width=0.7\textwidth]{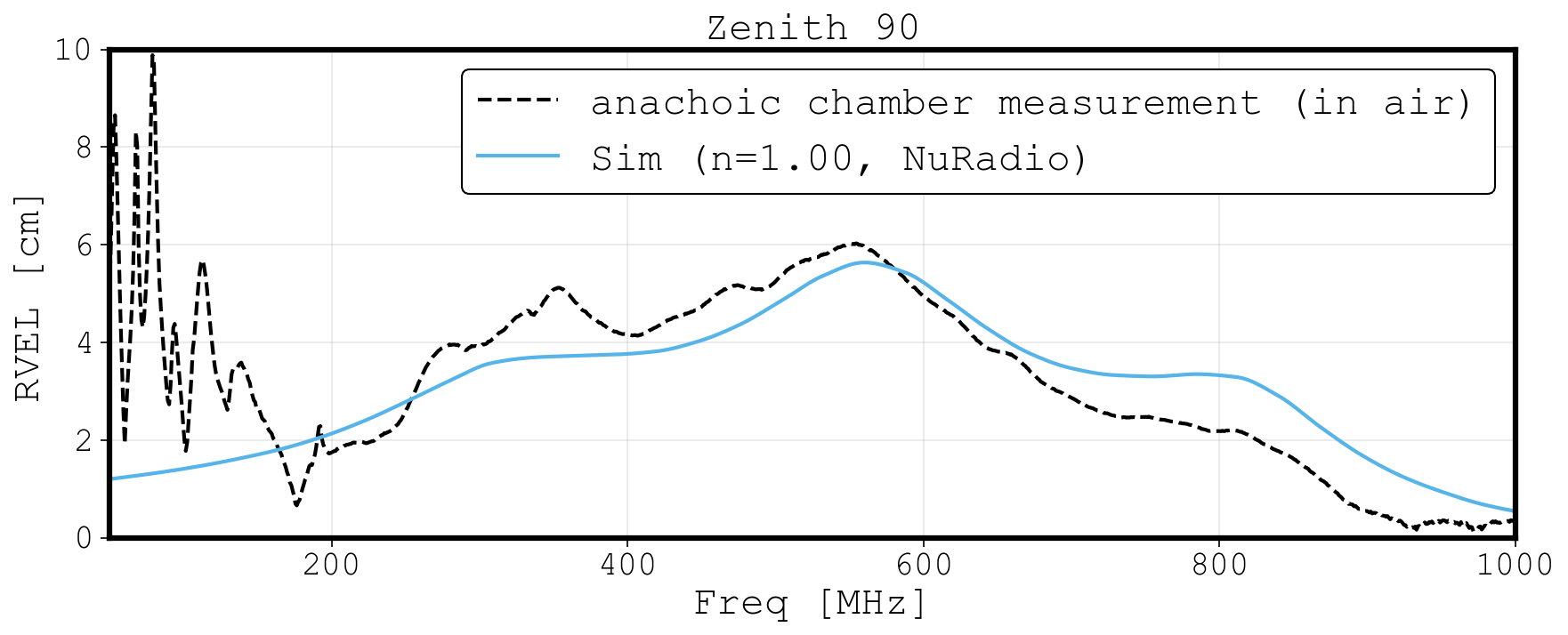}
    \caption{Same as Figure \ref{fig:rvel} but for an HPol antenna for the bore-sight direction.}
    \label{fig:rvel_2nd}
\end{figure}

\begin{figure}
	\centering
	\includegraphics[width=0.7\textwidth]{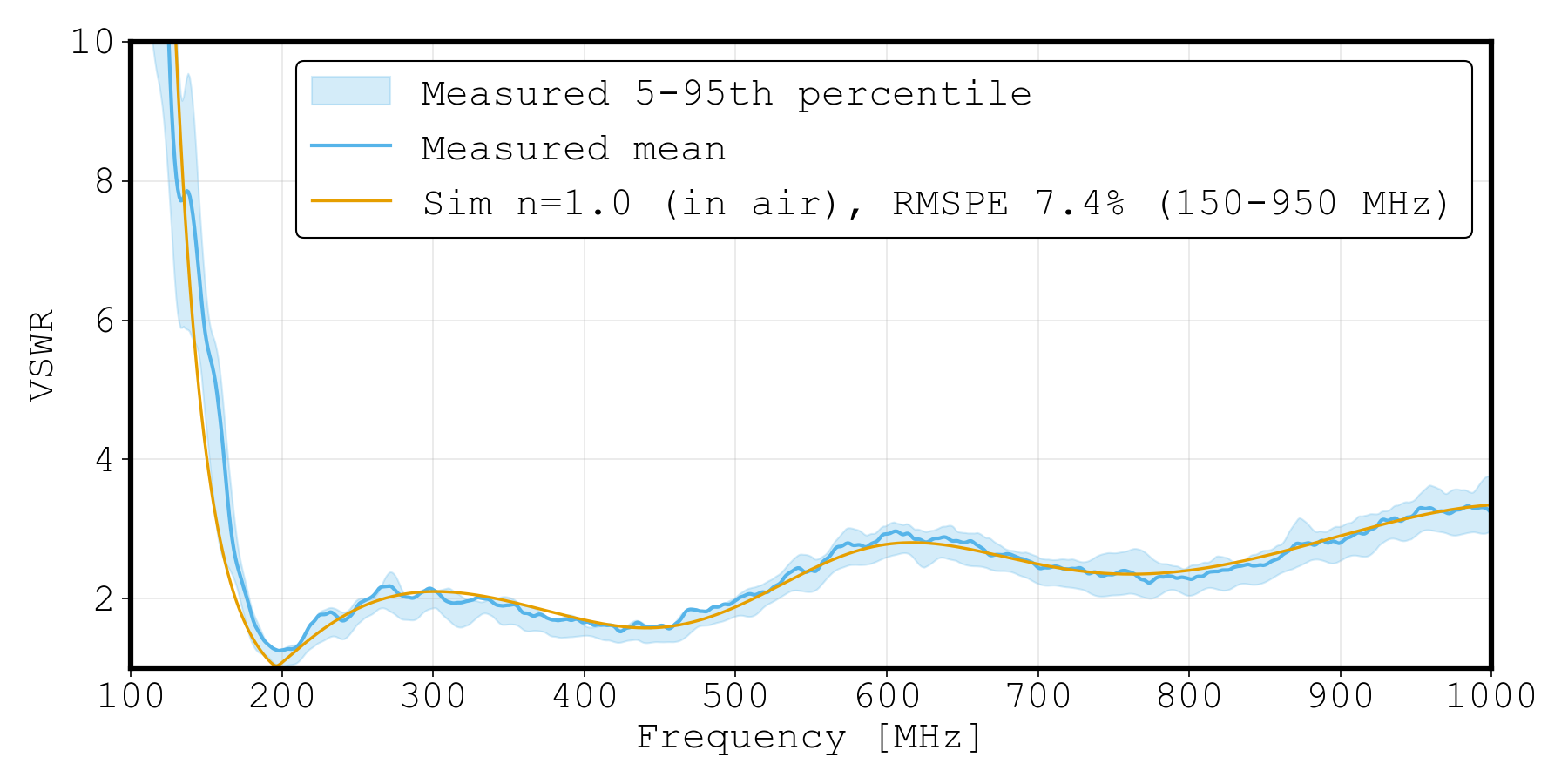}
	\includegraphics[width=0.7\textwidth]{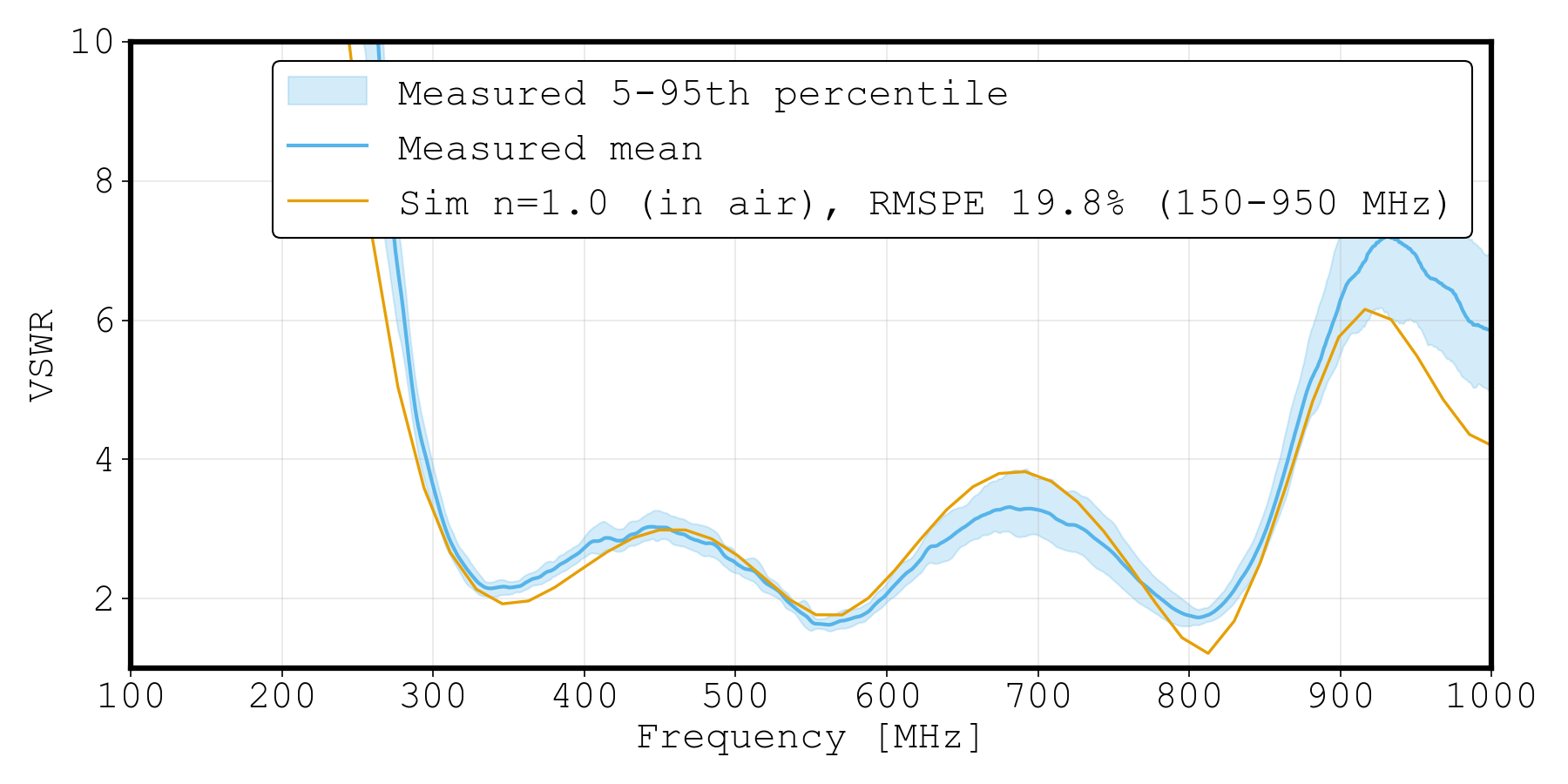}
	\caption{Voltage standing wave ratio (VSWR) for the VPol (top) and HPol (bottom) antennas. Good agreement over the relevant frequency range. RMSPE denotes the root mean squared percentage error.}
	\label{fig:vswr}
\end{figure}
\section{Electronic noise figures}
\label{app:electronic_noise_figure}

Section~\ref{sec:elec_noise} presents measurements of the electronic noise temperatures. Here, we describe how the noise temperature is derived from the measured noise factors. In particular, we explain how it is determined for the IGLUs, for which the lab setup differs from the in-situ configuration by the \SI{11}{m} coaxial cable connecting noise source and IGLU.

In our lab setup, a VNA measures the noise factor $F$ of several combined components. The noise factor of individual components relates to the total noise factor as
\begin{equation}
    \label{eq:noise_factor}
    F = F_1 + \frac{F_2 + 1}{G_1}
\end{equation}
where $F_i$ refers to the noise factor of components one and two and $G_1$ to the amplification of component one. Due to this relation, any noise sources after the first amplification stage, i.e., the SURFACE for the shallow component and the IGLU for the deep component, are negligible. For the shallow signal chain the measured noise figure can be directly interpreted as a noise temperature 
\begin{equation}
    T_\mathrm{ele} = (F - 1)T_0
\end{equation}
with the reference temperature $T_0 = 290$\,K. $T_\mathrm{ele}$ refers to the temperature at the antenna terminal (i.e., before amplification from the signal chain), which results in the measured noise power. For the deep component, we first have to subtract the effect of the 11\,m coaxial cable, which is present in the lab measurement (before the first amplification) but not in the deployed instrument. Using eq.~\ref{eq:noise_factor} with the coaxial cable (component 1) and the IGLU-fiber-DRAB triple (component 2) and further describing the cable's noise factor with $F_1 = 1 / G_1$ we get $F_2 = F\,G_1$. The cable gain is defined in power $G_1 = |S_{21}|^2$ with the scattering parameter $S_{21}$ measured by the VNA.

\end{document}